\documentclass[11pt]{article}
\usepackage{amssymb}
\usepackage{amsfonts}
\usepackage{amsmath}
\usepackage{geometry}
\usepackage[colorlinks=true, linkcolor=red, citecolor=blue]{hyperref}
\usepackage{float}
\usepackage{mathtools}
\usepackage{caption}
\usepackage{graphicx}
\usepackage[font=small,labelfont=bf,labelsep=space]{caption}
\usepackage{subfig}
\usepackage{setspace}

\makeatletter

\@addtoreset{equation}{section}
\makeatother

\input{tcilatex}
\begin{document}

\title{Generalized coherent states for the harmonic oscillator by the $J$%
-matrix method with an extension to the Morse potential}
\author{Hashim A. Yamani${}^{\ast }$ and Zouha\"{\i}r Mouayn$^{\diamond
,\natural }$ \\
$^{\ast }${\footnotesize \ Dar Al-Jewar, Knowledge Economic City, Medina,
Saudi Arabia\vspace*{-0.2em}}\\
{\footnotesize \ e-mail: hashim.haydara@gmail.com \vspace*{0.6mm}}\\
$^{\diamond }${\footnotesize \ Department of Mathematics, Faculty of
Sciences and Technics (M'Ghila),\vspace*{-0.2em}}\\
{\footnotesize \ Sultan Moulay Slimane University, BP. 523, B\'{e}ni Mellal
23000, Morocco }\\
$^{\natural }${\footnotesize Institut des Hautes \'{E}tudes Scientifiques,
Paris-Saclay University,}\\
{\footnotesize 35 route de Chartres, 91893 Bures-sur-Yvette, France.}\\
{\footnotesize e-mail: mouayn@gmail.com }\vspace*{-0.5em}}
\maketitle

\begin{abstract}
While dealing with the $J$-Matrix method for the harmonic oscillator to
write down its tridiagonal matrix representation in an orthonormal basis of $%
L^{2}\left( \mathbb{R}\right) ,$ we rederive a set of generalized coherent
states (GCS) of Perelomov type labeled by points $z$ of the complex plane $%
\mathbb{C}$ and depending on an integer number $m\in \mathbb{Z}_{+}$. The
number states expansion of these GCS gives rise to coefficients that are
complex Hermite polynomials whose linear superpositions provide
eigenfunctions for the two-dimensional magnetic Laplacian associated with
the $m$th Landau level. We extend this procedure to the Morse oscillator by
constructing a new set of GCS of Glauber type. 
\end{abstract}

\section{ Introduction}

Coherent states (CS) are mathematical tools which provide a close connection
between classical and quantum formalisms so as to play a central role in the
semiclassical analysis $\left[ 1\right] $. In general, CS are an
overcomplete family of normalized \textit{ket} vectors $\left\vert \zeta
\right\rangle $ which are labeled by points $\zeta $ of a phase-space domain 
$X$, belonging to a Hilbert space $\mathcal{H}$ that corresponds to a
specific quantum model and provides $\mathcal{H}$ with a resolution of its
identity operator with respect to a suitable integration measure on $X.$

In this paper, we will be dealing with the $J$-matrix method for the
harmonic oscillator Hamiltonian $H_{har}$ to write down its tridiagonal
matrix representation in an orthonormal basis of $\mathcal{H\equiv }%
L^{2}\left( \mathbb{R}\right) .$ We rederive a set of generalized coherent
states (GCS), say $\phi _{m}^{\omega ,z},$ of Perelomov type labeled by
points $z$ of the phase space $X\equiv \mathbb{C}$ and depending on an
integer number $m\in \mathbb{Z}_{+}$, obtained as orbits of the operators of
the Schr\"{o}dinger representation $T_{\omega }$, $\omega >0,$ of the
Heisenberg group on a $m$th Hermite function $h_{m}$. The resolution of the
identity operator on $L^{2}\left( \mathbb{R}\right) $ is ensured by the
square integrability of $T_{\omega }$ modulo the center of the group.
Beside, the number states expansion in the basis of eigenstates of $H_{har}$
gives rise to coefficients that are complex Hermite polynomials whose linear
superpositions provide eigenfunctions for the two-dimensional magnetic
Laplacian associated with the $m$th Landau level. This gives an alternative
answer to the problem rised by the authors $\left[ 2\right] $ who,
essentialy have used the cross-Wigner transform in order to intertwine
unitarly the two operators. Next we extend this procedure to the Morse
oscillator by constructing a new set of coherent states of Glauber type. 

\smallskip

The paper is organized as follows. In section 2 we review the classical
coherent states of the harmonic and some of their generalization depending
on the discret or continuous dynamics. In section 3, we review some
properties of a Hamiltonian having a tridiagonal representation in an
orthonormal basis of some Hilbert space. Section 4, deal with the
tridiagonalization of the harmonic oscillator which enables to recover a set
of generalized coherent states for this Hamiltonian. As application the
planar Landau problem and complex Hermite polynomials are also discussed. In
section 5 we give an extension of our procedure to the Hamiltonian with the
Morse potential.

\section{\noindent Coherent states}

\subsection{Canonical coherent states}

Coherent states were first introduced by E. Schr\"{o}dinger $\left[ 3\right] 
$ in order to obtain quantum states in $L^{2}(\mathbb{R})$ that follow the
classical flow associated to the harmonic oscillator Hamiltonian 
\begin{equation}
H_{har}=-\frac{\hbar }{2m_{\ast }}\frac{d^{2}}{d\xi ^{2}}+\frac{1}{2}\omega
^{2}\xi ^{2}-\frac{1}{2}\omega .  \tag{2.1}
\end{equation}%
Here, we consider the unit system where the Planck parameter $\hslash $, the
mass $m_{\ast }$ and $\omega $ equal one. These CS are ket vectors denoted
by $\left\vert z\right\rangle \in L^{2}(\mathbb{R})$ and labeled by elements
of $\,z\in $ $\mathbb{C}=\mathbb{R}^{2}$ which is the phase space of a
particle moving on $\mathbb{R}$, given by 
\begin{equation}
\langle \xi \left\vert z\right\rangle =\pi ^{-\frac{1}{4}}\left( e^{zz^{\ast
}}\right) ^{-\frac{1}{2}}\exp \left( -\frac{1}{2}z^{\ast 2}+\sqrt{2}z^{\ast
}\xi -\frac{1}{2}\xi ^{2}\right) ,\text{ \ \ }\xi \in \mathbb{R}.  \tag{2.2}
\end{equation}%
Their most important property is the resolution of the identity operator 
\begin{equation}
\mathbf{1}_{L^{2}(\mathbb{R})}=\frac{1}{\pi }\int\limits_{\mathbb{C}}d\nu
(z)\left\vert z\right\rangle \left\langle z\right\vert ,  \tag{2.3}
\end{equation}%
$d\nu (z)$ being the Lebesgue measure on $\mathbb{C}=\mathbb{R}^{2}.$ The
property $\left( 2.3\right) $ bridges between classical and quantum
mechanics in the sense that every operator acting on $L^{2}(\mathbb{R})$ or
any vector lying there may be decomposed over the phase space $\mathbb{C}$.
Moreover, the Bargmann transform $B_{0\text{ }}$defined by%
\begin{equation}
B_{0\text{ }}\left[ \varphi \right] \left( z\right) :=\left( e^{zz^{\ast
}}\right) ^{\frac{1}{2}}\langle \varphi \left\vert z\right\rangle _{L^{2}(%
\mathbb{R})}=\int\limits_{\mathbb{R}}\pi ^{-\frac{1}{4}}\exp \left( -\frac{1%
}{2}z^{\ast 2}+\sqrt{2}z^{\ast }\xi -\frac{1}{2}\xi ^{2}\right) \varphi
\left( \xi \right) d\xi ,\text{ \ \ }\varphi \in L^{2}(\mathbb{R})  \tag{2.4}
\end{equation}%
maps isometrically $L^{2}(\mathbb{R})$ onto the well known $\left[ 4\right] $
Fock-Bargmann space $\mathcal{F}\left( \mathbb{C}\right) $ of analytic
Gaussian square integrable functions on $\mathbb{C}$, whose reproducing
kernel is $K_{0}\left( z,w\right) $ $\propto e^{zw^{\ast }}$ for $z,w\in 
\mathbb{C}$. The CS in $\left( 2.2\right) $ may also be expanded over the
Fock basis of eigenfunctions $\left\vert \psi _{n}\right\rangle $ of the
operator $H_{har}$ as 
\begin{equation}
\left\vert z\right\rangle =\left( e^{zz^{\ast }}\right) ^{-\frac{1}{2}%
}\sum\limits_{n=0}^{+\infty }\left( \frac{z^{n}}{\sqrt{n!}}\right) ^{\ast
}\left\vert \psi _{n}\right\rangle .  \tag{2.5}
\end{equation}%
where the analytic coefficients $\left( \frac{1}{\sqrt{n!}}z^{n}\right) $ in
this expansion form an orthonormal basis of $\mathcal{F}\left( \mathbb{C}%
\right) $. This also means that the normalization prefactor in $\left(
2.5\right) $ encodes information on the diagonal of the reproducing kernel
function $K_{0}\left( z,z\right) .$

\subsection{\protect\smallskip Generalized coherent states}

For dynamics with discrete spectrum, one of the generalization of $\left(
2.5\right) $ to the finite dimensional case could described as follows. Let $%
(X,\sigma )$\ be a measure space and let $\mathcal{A}^{2}\subset
L^{2}(X,\sigma )$\ be a closed subspace of finite dimension $N$. Let $%
\left\{ D_{n}\right\} _{n=0}^{N-1}$ be an orthonormal system of functions in 
$\mathcal{A}^{2}$. For arbitrary $\zeta \in X,$ denote

\begin{equation}
\omega _{d}\left( \zeta \right) :=\sum_{n=0}^{N-1}\left\vert D_{n}\left(
\zeta \right) \right\vert ^{2}.  \tag{2.6}
\end{equation}%
Let $\mathfrak{H}_{d}$\ be a Hilbert space with dimension $N$ and $\left\{
\phi _{n}\right\} _{n=0}^{N-1}$\ be a complete orthonormal basis of $%
\mathfrak{H}_{d},$ which may be considered as eigenfunctions of some
Hamiltonian, associated with a finite spectrum $E_{0}<E_{1}<...<E_{N-1}$.
That is,\ 
\begin{equation}
1_{\mathfrak{H}_{d}}=\sum_{n=0}^{N-1}\left\vert \phi _{n}\right\rangle
\left\langle \phi _{n}\right\vert .  \tag{2.7}
\end{equation}%
\ The CS labeled by points $\zeta \in X$\ are defined as the ket-vectors $%
\left\vert \zeta \right\rangle \in \mathfrak{H}_{d}:$%
\begin{equation}
\left\vert \zeta \right\rangle _{d}:=\left( \omega _{d}\left( \zeta \right)
\right) ^{-\frac{1}{2}}\sum_{n=0}^{N-1}D_{n}\left( \zeta \right) \left\vert
\phi _{n}\right\rangle .\quad \quad  \tag{2.8}
\end{equation}%
Now, by definition $\left( 2.8\right) $, it is straightforward to show that $%
\langle u\left\vert u\right\rangle _{\mathfrak{H}_{d}}=1$\ and for $\phi
,\psi \in \mathfrak{H}_{d}$, we have 
\begin{equation}
\langle \phi \left\vert \psi \right\rangle _{\mathfrak{H}_{d}}=\int%
\limits_{X}d\sigma \left( \zeta \right) \omega _{d}\left( \zeta \right)
\langle \phi \left\vert \zeta \right\rangle _{\mathfrak{H}}\langle \zeta
\left\vert \psi \right\rangle _{\mathfrak{H}}.  \tag{2.9}
\end{equation}%
Thereby, we have a resolution of the identity operator of $\mathfrak{H}_{d}%
\mathfrak{,}$ which can be expressed as 
\begin{equation}
\mathbf{1}_{\mathfrak{H}_{d}}=\int\limits_{X}d\sigma \left( \zeta \right)
\omega _{d}\left( \zeta \right) \left\vert \zeta \right\rangle \left\langle
\zeta \right\vert \quad \quad  \tag{2.10}
\end{equation}%
where $\omega _{d}\left( \zeta \right) $\ appears as a weight function.

\smallskip

For the dynamics with continum spectra, we denote by $\left\vert \lambda
\right\rangle \in \mathcal{H}$ be an eigenfunction \ of the Hamiltonian $%
H\left\vert \lambda \right\rangle =E\left( \lambda \right) \left\vert
\lambda \right\rangle $ and for the continuous spectrum $\lambda $ may be
thought of some continuous variable such that the spectrum $E\left( \lambda
\right) $ is monotonically covered by variation of $\lambda $ from $\lambda
_{0}$ to $\infty .$ We assume that the continuous spectrum states are
normalized by the condition $\left\langle \lambda \right\vert \beta \rangle
\varpropto \delta \left( \lambda -\beta \right) .$ Consider a set of
functions $X\backepsilon \zeta \rightarrow C_{\lambda }\left( \zeta \right) $
\ indexed by $\lambda $, such that $\left\{ C_{\lambda }\right\} $ are
orthonormal in $L^{2}\left( X,d\mu \right) $%
\begin{equation}
\int\limits_{X}C_{\lambda }\left( \zeta \right) \overline{C_{\beta }\left(
\zeta \right) }d\mu \left( \zeta \right) =\delta \left( \lambda -\beta
\right) .  \tag{2.11}
\end{equation}%
Assume that 
\begin{equation}
\omega _{c}\left( \zeta \right) :=\int\limits_{\lambda _{0}}^{\infty
}\left\vert C_{\lambda }\left( \zeta \right) \right\vert ^{2}d\lambda <\infty
\tag{2.12}
\end{equation}%
The CS labeled by points $\zeta \in X$\ are defined as the ket-vectors $%
\left\vert \zeta \right\rangle \in \mathcal{H}$ by%
\begin{equation}
\left\vert \zeta \right\rangle _{c}:=\left( \omega _{c}\left( \zeta \right)
\right) ^{-\frac{1}{2}}\int\limits_{\lambda _{0}}^{\infty }C_{\lambda
}\left( \zeta \right) \left\vert \lambda \right\rangle d\lambda .\quad \quad
\tag{2.13}
\end{equation}%
We assume that there exists a measure $d\nu $ on $X$ such that 
\begin{equation}
\int\limits_{X}d\nu \left( \zeta \right) \omega _{c}\left( \zeta \right)
\left\vert \zeta \right\rangle \left\langle \zeta \right\vert
==\int\limits_{\lambda _{0}}^{\infty }\left\vert \lambda \right\rangle
\left\langle \lambda \right\vert d\lambda =1_{\mathcal{H}_{c}}  \tag{2.14}
\end{equation}%
One can define the generalized CS for both discrete and continuous spectra
via the combination 
\begin{equation}
\left\vert \zeta \right\rangle _{d,c}:=\left( \omega _{d}\left( \zeta
\right) \right) ^{-\frac{1}{2}}\sum_{n=0}^{N-1}D_{n}\left( \zeta \right)
\left\vert \phi _{n}\right\rangle +\left( \omega _{c}\left( \zeta \right)
\right) ^{-\frac{1}{2}}\int\limits_{\lambda _{0}}^{\infty }C_{\lambda
}\left( \zeta \right) \left\vert \lambda \right\rangle d\lambda .\quad 
\tag{2.15}
\end{equation}%
Of course we still have to precise the form of the suitable measure on $X$
that ensures the resolution of the identity operator which would be of the
form $1_{\mathcal{H}_{d}}+1_{\mathcal{H}_{c}}$, see $\left[ 5\right] .$
Actually, various generalizations of CS have been proposed. For an overview
of all aspects of the theory of CS and their genesis, we refer to the survey 
$\left[ 6\right] $ and $\left[ 1\right] ,\left[ 7\right] .$

\section{The $J$-matrix method}

We assume that we are given a positive semi-definite Hamiltonian $H$ (with
zero as the value of the lowest energy in its spectrum) acting on a Hilbert
space $\mathcal{H}$, that has a tridiagonal matrix representation in the
orthonormal basis $\{{\left\vert \phi _{n}\right\rangle }\}_{n=0}^{\infty }$
of $\mathcal{H}$ with known coefficients $\{a_{n},b_{n}\}_{n=0}^{\infty }$ 
\begin{equation}
H_{n,m}={\left\langle \phi _{n}\right\vert }H{\left\vert \phi
_{m}\right\rangle }=b_{n-1}\,\delta _{n,m+1}+\,a_{n}\,\delta
_{n,m}\,+\,b_{n}\,\delta _{n,m-1}.  \tag{3.1}
\end{equation}%
We solve the energy eigenvalue equation $H{\left\vert E\right\rangle }=E{%
\left\vert E\right\rangle }$ by expanding the eigenvector ${\left\vert
E\right\rangle }$ in the basis ${\left\vert \phi _{n}\right\rangle }$ as ${%
\left\vert E\right\rangle }=\sum_{n=0}^{\infty }f_{n}(E){\left\vert \phi
_{n}\right\rangle }$. Making use of the tridiagonality of $H$, we readily
obtain the following recurrence relations for the expansion coefficients: 
\begin{equation}
\begin{array}{l}
{Ef_{0}(E)=a_{0}f_{0}(E)+b_{0}f_{1}(E)\quad \quad } \\ 
{Ef_{n}(E)=b_{n-1}f_{n-1}(E)+a_{n}f_{n}(E)+b_{n}f_{n+1}(E),\quad
n=1,2,...,.\quad }%
\end{array}
\tag{3.2}
\end{equation}%
The operator $H$ may admit a continuous spectrum $\sigma _{c}$ and a
discrete part $\{E_{\mu }\}_{\mu }$, both of which lead to the following
form of the resolution of the identity operator on $\mathcal{H}$: 
\begin{equation}
\sum_{\mu }{\left\vert E_{\mu }\right\rangle }{\left\langle E_{\mu
}\right\vert }+\int_{\sigma _{c}}{\left\vert E\right\rangle }{\left\langle
E\right\vert }\,dE\;=1_{\mathcal{H}}.\quad \quad  \tag{3.3}
\end{equation}%
This translates into the following orthogonality relation for the expansion
coefficients: 
\begin{equation}
\sum_{\mu }f_{n}(E_{\mu })\left( f_{m}(E_{\mu })\right) ^{\ast
}+\int_{\sigma _{c}}f_{n}(E)\left( f_{m}(E)\right) ^{\ast }dE\,=\delta
_{n,m}.  \tag{3.4}
\end{equation}%
If we now define $p_{n}(E)=\frac{f_{n}(E)}{f_{0}(E)}$, then $\{p_{n}(E)\}$
is a set of polynomials that satisfy the three-term recursion relation 
\begin{equation}
Ep_{n}(E)=b_{n-1}p_{n-1}(E)+a_{n}p_{n}(E)+b_{n}p_{n+1}(E),\quad n=1,2,..., 
\tag{3.5}
\end{equation}%
with the initial conditions $p_{0}(E)=1$ and $p_{1}(E)=(E-a_{0})b_{0}^{-1}$.
If we further define $\Omega (E)=\left\vert f_{0}(E)\right\vert ^{2}$ and $%
\Omega _{\mu }=\left\vert f_{0}(E_{\mu })\right\vert ^{2}$, then the
relation $\left( 2.4\right) $ now translates into the following
orthogonality relation for the polynomial $p_{n}$: 
\begin{equation}
\sum_{\mu }\Omega _{\mu }p_{n}(E_{\mu })\left( p_{m}(E_{\mu })\right) ^{\ast
}+\int\limits_{\sigma _{c}}\Omega (E)p_{n}(E)\left( p_{m}\left( E\right)
\right) ^{\ast }dE\,=\delta _{n,m}.  \tag{3.6}
\end{equation}

We have shown $\left[ 8\right] $ that we can write the Hamiltonian $H$ in
the form $H=A^{\dag }A$, where the forward-shift operator $A$ is defined by
its action on the basis vector as 
\begin{equation}
A{\left\vert \phi _{n}\right\rangle }=c_{n}\,{\left\vert \phi
_{n}\right\rangle }+\,d_{n}\,{\left\vert \phi _{n-1}\right\rangle }. 
\tag{3.7}
\end{equation}%
Furtheremore, we require from the adjoint operator $A^{\dag }$ to act on the
ket vectors ${\left\vert \phi _{n}\right\rangle }$ in the following way 
\begin{equation}
A^{\dagger }{\left\vert \phi _{n}\right\rangle }=c_{n}^{\ast }\,{\left\vert
\phi _{n}\right\rangle }+\,d_{n+1}^{\ast }\,{\left\vert \phi
_{n+1}\right\rangle .}  \tag{3.8}
\end{equation}%
Here, the coefficients $\{c_{n},d_{n}\}_{n=0}^{\infty }$ are related to the
coefficients $\{a_{n},b_{n}\}_{n=0}^{\infty }$ and the polynomials $%
\{p_{n}\}_{n=0}^{\infty }$ 
\begin{equation}
{d_{0}=0,\,\;c_{n}}c_{n}^{\ast }{=-b_{n}\frac{p_{n+1}(0)}{p_{n}(0)},\quad
d_{n+1}d_{n+1}^{\ast }=-b_{n}\frac{p_{n}(0)}{p_{n+1}(0)},\quad }{a_{n}=c}%
_{n}c_{n}^{\ast }+{d_{n}d_{n}^{\ast },\quad b_{n}=c_{n}d_{n+1}^{\ast }}. 
\tag{3.9}
\end{equation}

\section{Tridiagonalization of the harmonic oscillator}

\subsection{Construction of the basis $\left( \Phi _{m}^{z,\protect\omega %
}\right) _{m\geq 0}$}

We recall the following regarding the one-dimensional harmonic oscillator

\begin{equation}
H_{har}=-\frac{1}{2}\frac{d^{2}}{d\xi ^{2}}+\frac{1}{2}\omega ^{2}\xi ^{2}-%
\frac{1}{2}\omega ,  \tag{4.1}
\end{equation}%
$\omega >0$ being a fixed parameter. It decomposes as 
\begin{equation}
H_{har}=A^{\dagger }A  \tag{4.2}
\end{equation}%
where 
\begin{equation}
A=-\frac{1}{\sqrt{2}}\frac{d}{d\xi }+W\left( \xi \right) ,\text{ \ }%
A^{\dagger }=\frac{1}{\sqrt{2}}\frac{d}{d\xi }+W\left( \xi \right) ,\text{ \ 
}W\left( \xi \right) =-\frac{\omega }{\sqrt{2}}\xi .  \tag{4.3}
\end{equation}%
The spectrum of $H_{har}$ in $L^{2}\left( \mathbb{R},d\xi \right) $ consists
on a set of discrete eigenvalues $\varepsilon _{n}=n\omega ,$ $n=0,1,2,...,$
with the corresponding eigenfunctions given by 
\begin{equation}
\psi _{n}\left( \xi \right) =\left( -1\right) ^{n}\left( \frac{\omega }{\pi }%
\right) ^{\frac{1}{4}}\left( 2^{n}n!\right) ^{-\frac{1}{2}}e^{-\frac{1}{2}%
\omega \xi ^{2}}H_{n}\left( \sqrt{\omega }\xi \right) ,\text{ \ \ }\xi \in 
\mathbb{R},  \tag{4.4}
\end{equation}%
in terms of the Hermite polynomial $\left( \left[ 9\right] \text{, p.100}%
\right) :$

\begin{equation}
H_{n}\left( u\right) =n!\sum\limits_{\ell =0}^{\left\lfloor \frac{1}{2}%
n\right\rfloor }\frac{\left( -1\right) ^{\ell }\left( 2u\right) ^{n-2\ell }}{%
\ell !\left( n-2\ell \right) !},\text{ \ \ }u\in \mathbb{R}.  \tag{4.5}
\end{equation}%
But and since the Hamiltonian is shape invariant, we know $\left[ 8\right] $
that we can generate the infinite set of coefficients $\left\{
c_{n},d_{n}\right\} $ from the three values, $(d_{0},d_{1},c_{0})$ together
with the energy spectrum of the system, as follows 
\begin{equation}
c_{n}^{2}=c_{0}^{2}+\left[ nd_{1}^{2}-\varepsilon _{n}\right] ,\quad
d_{n}^{2}=nd_{1}^{2}+\left[ n\varepsilon _{1}-\varepsilon _{n}\right] . 
\tag{4.6}
\end{equation}%
Starting now with $d_{0}=0,$ $d_{1}=\sqrt{\omega }$ and $c_{0}=-z^{\ast },$ $%
z\in \mathbb{C}$. \ It is easy to deduce that $\left( 4.6\right) $ gives the
simple results $c_{n}^{2}=z^{\ast 2}$ and $d_{n}^{2}=n\omega $. Therefore,
the coefficients of the Hamiltonian tridiagonal matrix are of the form

\begin{equation}
a_{n}\equiv a_{n}\left( z,\omega \right) =\left\vert c_{n}\right\vert
^{2}+\left\vert d_{n}\right\vert ^{2}=zz^{\ast }+n\omega ,  \tag{4.7}
\end{equation}

\begin{equation}
b_{n}\equiv b_{n}\left( z,\omega \right) =c_{n}d_{n+1}^{\ast }=-z^{\ast }%
\sqrt{(n+1)\omega }.  \tag{4.8}
\end{equation}%
This is a situation where we know the coefficients $\left\{
a_{n},b_{n}\right\} $ explicitly, but we do not know the elements of the
tridiagonal basis, say $\left\{ \Phi _{n}^{z,\omega }\right\} _{n\geq 0}$,
except for the first two elements. Indeed, we start with%
\begin{equation*}
\Phi _{0}^{z,\omega }\left( \xi \right) =\left( \frac{\omega }{\pi }\right)
^{\frac{1}{4}}\left( e^{\frac{1}{\omega }zz^{\ast }}\right) ^{-\frac{1}{2}%
}\exp \left( -\frac{1}{2\omega }z^{\ast 2}+\sqrt{2}z^{\ast }\xi -\frac{1}{2}%
\omega \xi ^{2}\right)
\end{equation*}%
which is the canonical CS given in $\left( 2.2\right) .$ From the relations
in $\left( 4.3\right) ,$ we have $A+$ $A^{\dagger }=2W$. \ Hence, $%
A^{\dagger }=W-A.$ Since $A\Phi _{0}^{z,\omega }=z\Phi _{0}^{z,\omega }$and $%
A^{\dagger }\Phi _{0}^{z,\omega }=z^{\ast }\Phi _{0}^{z,\omega }+d_{1}\Phi
_{1}^{z,\omega },$we deduce that $d_{1}\Phi _{1}^{z,\omega }=2\left( W-\func{%
Re}z\right) \Phi _{1}^{z,\omega }.$ Thus,%
\begin{equation*}
d_{1}d_{1}^{\ast }=4\int\limits_{\mathbb{R}}\left( W\left( \xi \right) -%
\func{Re}z\right) ^{2}\left\vert \Phi _{0}^{z,\omega }\left( \xi \right)
\right\vert ^{2}d\xi =\omega .
\end{equation*}%
Now, with this explicit value of $d_{1},$ we obtain $\Phi _{1}^{z,\omega }$
explicitly as 
\begin{equation*}
\Phi _{1}^{z,\omega }=\left( \frac{\omega }{\pi }\right) ^{\frac{1}{4}}\frac{%
-1}{\sqrt{2}}\left( e^{\frac{1}{\omega }zz^{\ast }}\right) ^{-\frac{1}{2}%
}\exp \left( -\frac{1}{2\omega }z^{\ast 2}+\sqrt{2}z^{\ast }\xi -\frac{1}{2}%
\omega \xi ^{2}\right) H_{1}\left( \sqrt{\omega }\xi -\sqrt{\frac{2}{\omega }%
}\func{Re}z\right) .
\end{equation*}%
To determine the full basis elements, we first proceed by fixing an
eigenfunction $\psi _{s}$ in $\left( 4.4\right) $ and expanding it over
elements of the basis $\left\{ \Phi _{n}^{z,\omega }\right\} _{n\geq 0}$ as%
\begin{equation}
\psi _{s}(\xi )=\sum_{n=0}^{\infty }\left[ \varrho \left( \varepsilon
_{s}\right) p_{n}\left( \varepsilon _{s}\right) \right] \Phi _{n}^{z,\omega
}\left( \xi \right) ,\text{ \ }\xi \in \mathbb{R}  \tag{4.9}
\end{equation}%
with $p_{n}$ being an orthogonal polynomial satisfying the conditions 
\begin{equation}
\sum_{s=0}^{\infty }\varrho \left( \varepsilon _{s}\right) p_{n}\left(
\varepsilon _{s}\right) \left( \varrho \left( \varepsilon _{s}\right)
p_{m}\left( \varepsilon _{s}\right) \right) ^{\ast }=\delta _{n,m},\text{ \ }%
p_{0}\left( \varepsilon _{s}\right) =1.  \tag{4.10}
\end{equation}%
where $\varrho \left( \bullet \right) $ stands for the square root of an
orthonormalization density function. From the tridiagonality condition, we
know that $p_{n}$ satisfy the following three-term recurrence relation 
\begin{equation}
\varepsilon _{s}p_{n}\left( \varepsilon _{s}\right) =b_{n-1}p_{n-1}\left(
\varepsilon _{s}\right) +a_{n}p_{n}\left( \varepsilon _{s}\right)
+b_{n}^{\ast }p_{n+1}\left( \varepsilon _{s}\right) .  \tag{4.11}
\end{equation}%
By $\left( 4.7\right) $-$\left( 4.8\right) $, Eq.$\left( 4.11\right) $ takes
the form 
\begin{equation}
\varepsilon _{s}p_{n}\left( \varepsilon _{s}\right) =-z^{\ast }\sqrt{n\omega 
}p_{n-1}\left( \varepsilon _{s}\right) +\left( \left\vert z\right\vert
^{2}+n\omega \right) p_{n}\left( \varepsilon _{s}\right) -z\sqrt{\left(
n+1\right) \omega }p_{n+1}\left( \varepsilon _{s}\right) .  \tag{4.12}
\end{equation}%
We introduce the follwing change of polynomial by setting 
\begin{equation}
p_{n}\left( u\right) =\omega ^{-\frac{1}{2}n}\frac{z^{\ast n}}{\sqrt{n!}}%
Q_{n}\left( u;\left\vert z\right\vert ^{2}\right) ,\text{ \ }u\equiv
\varepsilon _{s}  \tag{4.13}
\end{equation}%
where $u\mapsto Q_{n}\left( u;\left\vert z\right\vert ^{2}\right) $ is a
polynomial, depending on the parameter $\left\vert z\right\vert
^{2}=zz^{\ast }$, to be determined. For this we insert $\left( 4.13\right) $
into $\left( 4.12\right) ,$ to obtain the followng equation

\begin{equation}
-\varepsilon _{s}Q_{n}\left( \varepsilon _{s};\left\vert z\right\vert
^{2}\right) =zz^{\ast }Q_{n+1}\left( \varepsilon _{s};\left\vert
z\right\vert ^{2}\right) -\left( \left\vert z\right\vert ^{2}+n\omega
\right) Q_{n}\left( \varepsilon _{s};\left\vert z\right\vert ^{2}\right)
+n\omega Q_{n-1}\left( \varepsilon _{s};\left\vert z\right\vert ^{2}\right) ,%
\text{ \ }\varepsilon _{s}=s\omega .  \tag{4.14}
\end{equation}%
Inroducing the function%
\begin{equation}
\mathbb{N}\ni s\mapsto g_{n}^{\left( \omega ,\left\vert z\right\vert
^{2}\right) }\left( s\right) :=Q_{n}\left( s\omega ;\left\vert z\right\vert
^{2}\right) ,  \tag{4.15}
\end{equation}%
then, eq. $\left( 4.14\right) $ transforms to%
\begin{equation}
-sg_{n}^{\left( \omega ,\left\vert z\right\vert ^{2}\right) }\left( s\right)
=ag_{n+1}^{\left( \omega ,\left\vert z\right\vert ^{2}\right) }\left(
s\right) -\left( n+a\right) g_{n}^{\left( \omega ,\left\vert z\right\vert
^{2}\right) }\left( s\right) +ng_{n-1}^{\left( \omega ,\left\vert
z\right\vert ^{2}\right) }\left( s\right) ,\text{ \ \ }a=\omega
^{-1}zz^{\ast }.  \tag{4.16}
\end{equation}%
From the latter one we recognize the three-term recurrence relation of the
Charlier polynomial ($\left[ 10\right] ):$

\begin{equation}
-uC_{n}\left( u;a\right) =aC_{n+1}\left( u;a\right) -\left( n+a\right)
C_{n}\left( u;a\right) +nC_{n-1}\left( u;a\right) ,\text{ \ \ }a>0 
\tag{4.17}
\end{equation}%
which may also be defined in terms of the Laguerre polynomial as 
\begin{equation}
u\mapsto C_{n}\left( u;a\right) =\frac{n!}{\left( -a\right) ^{n}}%
L_{n}^{\left( u-n\right) }\left( a\right) =\frac{1}{\left( -a\right) ^{n}}%
\sum\limits_{k=0}^{n}\frac{\left( -n\right) _{k}}{k!}\left( u-n+k+1\right)
_{n-k}a^{k}  \tag{4.18}
\end{equation}%
Thus, 
\begin{equation}
g_{n}^{\left( \omega ,\left\vert z\right\vert ^{2}\right) }\left( s\right)
=\left( -1\right) ^{n}n!\left( \frac{\omega }{\left\vert z\right\vert ^{2}}%
\right) ^{n}L_{n}^{\left( s-n\right) }\left( \frac{\left\vert z\right\vert
^{2}}{\omega }\right)  \tag{4.19}
\end{equation}%
Consequently,%
\begin{equation}
p_{n}\left( \varepsilon _{s}\right) =\left( -1\right) ^{n}\sqrt{n!}\left( 
\sqrt{\omega }\right) ^{n}z^{-n}L_{n}^{\left( s-n\right) }\left( \frac{1}{%
\omega }zz^{\ast }\right)  \tag{4.20}
\end{equation}%
From the orthogonality relations of the Charlier polynomials $:$%
\begin{equation}
\sum\limits_{u=0}^{\infty }\frac{a^{u}}{u!}C_{n}\left( u;a\right)
C_{m}^{\ast }\left( u;a\right) =a^{-n}e^{a}n!\delta _{n,m}  \tag{4.22}
\end{equation}%
we deduce that%
\begin{equation}
\sum\limits_{s=0}^{\infty }\left( \frac{1}{\sqrt{s!}}\left( \frac{z}{\sqrt{%
\omega }}\right) ^{s}e^{-\frac{1}{2}\frac{zz^{\ast }}{\omega }}p_{n}\left(
\varepsilon _{s}\right) \right) \left( \frac{1}{\sqrt{s!}}\left( \frac{z}{%
\sqrt{\omega }}\right) ^{s}e^{-\frac{1}{2}\frac{zz^{\ast }}{\omega }%
}p_{m}\left( \varepsilon _{s}\right) \right) ^{\ast }=\delta _{n,m}, 
\tag{4.23}
\end{equation}%
meaning that the unknown coefficients $\left( \varrho \left( \varepsilon
_{s}\right) p_{n}\left( \varepsilon _{s}\right) \right) _{n\geq 0}$ in the
expansion $\left( 4.9\right) $ are of the form%
\begin{equation}
\varrho \left( \varepsilon _{s}\right) p_{n}\left( \varepsilon _{s}\right) =%
\frac{1}{\sqrt{s!}}\left( \frac{z}{\sqrt{\omega }}\right) ^{s}e^{-\frac{1}{2}%
\frac{zz^{\ast }}{\omega }}p_{n}\left( \varepsilon _{s}\right) .  \tag{4.24}
\end{equation}%
Using the orthognality relation $\left( 4.23\right) $, we can invert the
relation $\left( 4.9\right) $ in order to expand a fixed vector $\Phi
_{m}^{z,\omega }$ over the eigenvectors basis of Hermite functions $\left(
\psi _{s}\right) _{s\geq 0}$ as follows

\begin{equation}
\Phi _{m}^{z,\omega }=\sum_{s=0}^{\infty }\left[ \varrho \left( \varepsilon
_{s}\right) p_{m}\left( \varepsilon _{s}\right) \right] ^{\ast }\psi _{s}. 
\tag{4.25}
\end{equation}%
Explicitly, in the $\xi $-coordinate, we have 
\begin{equation}
\Phi _{m}^{z,\omega }\left( \xi \right) =\left( -1\right) ^{m}\sqrt{m!}%
\left( \sqrt{\omega }\right) ^{m}\frac{1}{z^{\ast m}}e^{-\frac{1}{2}\frac{%
zz^{\ast }}{\omega }}\sum_{s=0}^{\infty }\left[ \frac{1}{\sqrt{s!}}\frac{%
z^{\ast s}}{\sqrt{\omega }^{s}}L_{m}^{\left( s-m\right) }\left( \frac{1}{%
\omega }zz^{\ast }\right) \right] \psi _{s}\left( \xi \right) .  \tag{4.26}
\end{equation}%
We replace $\psi _{s}\left( \xi \right) $ by its expession $\left(
4.4\right) $ and we prove (see Appendix A) that the sum in the r.h.s of $%
\left( 4.26\right) $ can be written in a closed form as%
\begin{equation}
\left( \frac{\omega }{\pi }\right) ^{\frac{1}{4}}e^{\frac{1}{2}\omega \xi
^{2}}\frac{1}{m!}z^{\ast m}\left( \frac{1}{\sqrt{2\omega }}\right) ^{m}\exp
\left( \left( \frac{iz^{\ast }}{\sqrt{2\omega }}-i\sqrt{\omega }\xi \right)
^{2}\right) H_{m}\left( \sqrt{\omega }\xi -\sqrt{\frac{2}{\omega }}\func{Re}%
z\right) .  \tag{4.27}
\end{equation}%
Therefore, Eq.$\left( 4.26\right) $ reads

\begin{equation}
\Phi _{m}^{z,\omega }\left( \xi \right) =\left( \frac{\omega }{\pi }\right)
^{\frac{1}{4}}\frac{\left( -1\right) ^{m}}{\sqrt{m!2^{m}}}\left( e^{\frac{1}{%
\omega }zz^{\ast }}\right) ^{-\frac{1}{2}}\exp \left( -\frac{1}{2\omega }%
z^{\ast 2}+\sqrt{2}z^{\ast }\xi -\frac{1}{2}\omega \xi ^{2}\right)
H_{m}\left( \sqrt{\omega }\xi -\sqrt{\frac{2}{\omega }}\func{Re}z\right) 
\tag{4.28}
\end{equation}%
To check that the basis set $\left\{ \Phi _{m}^{z,\omega }\right\} _{m\geq
0} $ does indeed tridiagonalize the one-dimensional harmonic oscillator we
may use the derivative of the Hermite polynomial 
\begin{equation}
\frac{d}{du}H_{n}(u)=2nH_{n-1}(u)  \tag{4.29}
\end{equation}%
to obtain that%
\begin{equation}
\frac{d}{d\xi }\Phi _{m}^{z,\omega }=\sqrt{2}z^{\ast }\Phi _{m}^{z,\omega
}+2m\sqrt{\omega }\Phi _{m-1}^{z,\omega }-\omega \xi \Phi _{m}^{z,\omega } 
\tag{4.30}
\end{equation}%
which can be used to see that the operator $A=-\frac{1}{\sqrt{2}}\frac{d}{%
d\xi }-\frac{1}{\sqrt{2}}\omega \xi $ acts on $\Phi _{m}^{z,\omega }$ as 
\begin{equation}
A\Phi _{m}^{z,\omega }=-z^{\ast }\Phi _{m}^{z,\omega }+\sqrt{m\omega }\Phi
_{m-1}^{z,\omega }.  \tag{4.31}
\end{equation}%
Next, by applying to the last term $\xi \Phi _{m}^{z,\omega }$ in the r.h.s
of $\left( 4.30\right) $ the three-term recurrence relation of the Hermite
polynomial 
\begin{equation}
2uH_{n}(u)=H_{n+1}(u)+2nH_{n-1}(u)  \tag{4.32}
\end{equation}%
direct calculations show that the operator $A^{\dagger }=\frac{1}{\sqrt{2}}%
\frac{d}{d\xi }-\frac{1}{\sqrt{2}}\omega \xi $ acts on $\Phi _{m}^{z,\omega }
$ as 
\begin{equation}
A^{\dagger }\Phi _{m}^{z,\omega }=-z\Phi _{m}^{z,\omega }+\sqrt{\left(
m+1\right) \omega }\Phi _{m+1}^{z,\omega }.  \tag{4.33}
\end{equation}%
Note that $\left( 4.31\right) $ and $\left( 4.32\right) $ were expected in
view of the general setting $\left( 3.7\right) -\left( 3.8\right) $ together
with the choice of $c_{n}=-z^{\ast }$ and $d_{n}=\sqrt{n\omega }.$ Finally,
By using the factorization $H_{har}=A^{\dagger }A$ and the above equations $%
\left( 4.31\right) $, $\left( 4.32\right) ,$ we arrive at the tridiagonal
form

\begin{equation}
\left\langle \Phi _{n}^{z,\omega }|H_{har}|\Phi _{m}^{z,\omega
}\right\rangle =-z\sqrt{n\omega }\delta _{n,m+1}+\left( n\omega +zz^{\ast
}\right) \delta _{nm}-z^{\ast }\sqrt{(n+1)\omega }\delta _{n,m-1}  \tag{4.34}
\end{equation}%
We may use $\left( 4.31\right) ,\left( 4.32\right) $ and $\left( 4.34\right) 
$ to check that the expectations of the operator $H_{har}$ and its square $%
H_{har}^{2}$ with respect to the initial basis vector $\Phi _{0}^{z,\omega }$
are respectively given by $\left\langle \Phi _{0}^{z,\omega }|H_{har}|\Phi
_{0}^{z,\omega }\right\rangle =zz^{\ast }$ and $\left\langle \Phi
_{0}^{z,\omega }|H_{har}^{2}|\Phi _{0}^{z,\omega }\right\rangle =zz^{\ast
}\left( \omega +zz^{\ast }\right) $. As a matter of fact, the coefficient $%
c_{0}$ and $d_{1}$ which have been fixed as a starting step in our procedure
to find out the rest of the basis vectors $\Phi _{1}^{z,\omega },\Phi
_{2}^{z,\omega },\Phi _{3}^{z,\omega },...$, can also be determined by the
above expectation as follows%
\begin{equation}
c_{0}c_{0}^{\ast }=\mathbb{E}_{\Phi _{0}^{z,\omega }}\left( H_{har}\right)
=zz^{\ast }  \tag{4.35}
\end{equation}%
and 
\begin{equation}
d_{1}d_{1}^{\ast }=\frac{\mathbb{V}_{\Phi _{0}^{z,\omega }}\left(
H_{har}\right) }{\mathbb{E}_{\Phi _{0}^{z,\omega }}\left( H_{har}\right) }=%
\frac{\left\langle \Phi _{0}^{z,\omega }|H_{har}^{2}|\Phi _{0}^{z,\omega
}\right\rangle }{\left\langle \Phi _{0}^{z,\omega }|H_{har}|\Phi
_{0}^{z,\omega }\right\rangle }-\left\langle \Phi _{0}^{z,\omega
}|H_{har}|\Phi _{0}^{z,\omega }\right\rangle =\omega .  \tag{4.36}
\end{equation}%
\textbf{Remark 4. 1. }Observe that $m$ $\in \mathbb{Z}_{+}$ is fixed in the
series\textbf{\ }%
\begin{equation}
\Phi _{m}^{z,\omega }=\sum_{s=0}^{\infty }\left[ \varrho \left( \varepsilon
_{s}\right) p_{m}\left( \varepsilon _{s}\right) \right] ^{\ast }\psi
_{s}=\left( e^{\frac{1}{\omega }zz^{\ast }}\right) ^{-\frac{1}{2}%
}\sum_{s=0}^{\infty }\left[ C_{s}^{\left( m,\omega \right) }\left( z,z^{\ast
}\right) \right] ^{\ast }\psi _{s}  \tag{4.37}
\end{equation}%
while $s$ is runing over all positive integers. Therefore, we may use the
identity on the Laguerre polynomial $\left( \left[ 11\right] \text{, p.98}%
\right) :$ 
\begin{equation}
L_{n}^{\left( -r\right) }\left( t\right) =\left( -t\right) ^{r}\frac{\left(
n-r\right) !}{n!}L_{n-r}^{\left( r\right) }\left( t\right) ,\text{ \ \ }%
1\leq r\leq n  \tag{4.38}
\end{equation}%
to present the coefficients in $\left( 4.35\right) $ in the form%
\begin{equation}
C_{s}^{\left( m,\omega \right) }\left( z,z^{\ast }\right) =\frac{1}{\sqrt{s!}%
}\left( \frac{z}{\sqrt{\omega }}\right) ^{s}p_{m}\left( \varepsilon
_{s}\right) =\left( -1\right) ^{m\wedge s}\frac{\sqrt{\left( m\wedge
s\right) !}}{\sqrt{\left( m\vee s\right) !}}\left( \frac{zz^{\ast }}{\omega }%
\right) ^{\frac{1}{2}\left\vert m-s\right\vert }e^{i\left( s-m\right) \theta
}L_{m\wedge s}^{\left\vert m-s\right\vert }\left( \frac{zz^{\ast }}{\omega }%
\right) ,\text{ \ }\theta =\arg z  \tag{4.39}
\end{equation}%
which, in fact, coincide with the so called complex Hermite polynomials 
\begin{equation}
C_{s}^{\left( m,\omega \right) }\left( z,z^{\ast }\right) =H_{m,s}\left( 
\frac{z}{\sqrt{\omega }},\frac{z^{\ast }}{\sqrt{\omega }}\right)
=\sum\limits_{j=0}^{m\wedge s}\left( -1\right) ^{j}\frac{m!s!}{j!\left(
m-j\right) !\left( s-j\right) !}\left( \frac{z}{\sqrt{\omega }}\right)
^{m-j}\left( \frac{z^{\ast }}{\sqrt{\omega }}\right) ^{s-j}  \tag{4.40}
\end{equation}%
which were first introduced by It\^{o} $\left[ 12\right] ;$ see $\left[ 13%
\right] $ and references therein.

\subsection{$\left( \Phi _{m}^{z,\protect\omega }\right) _{z\in \mathbb{C}}$
as a set Perelomov's coherent states for the harmonic oscillator}

The Heisenberg group $\mathbb{H}_{1}$ (of degree $1$) is the Lie group whose
underlying manifold is $\mathbb{C\times R=R}^{3}$ with coordinates ($q,p,t)$%
\ and whose group law is $(q,p,t).(q^{\prime },p^{\prime },t^{\prime
})=(q+q^{\prime },p+p^{\prime },t+t^{\prime }+\frac{1}{2}(qp^{^{\prime
}}-q^{\prime }p)).$ The continuous unitary irreducible representations (UIR)
of $\mathbb{H}_{1}$ are well known ($\left[ 14\right] $, p$.$37) Here, we
will be concerned with a UIR\ of $\mathbb{H}_{1}$ on the Hilbert space $%
L^{2}(\mathbb{R},d\xi )$ by shift and multiplication operators $\ $($\left[
15\right] $, \S 1.1): 
\begin{equation}
T_{\omega }(q,p,t)\left[ \psi \right] (\xi ):=\exp i\left( \frac{2}{\omega
^{2}}t-\frac{\sqrt{2}}{\omega }p\xi +\frac{1}{\omega ^{2}}qp\right) \psi
\left( \xi -\frac{\sqrt{2}}{\omega }q\right) ,\xi \in \mathbb{R}\text{,} 
\tag{4.39}
\end{equation}%
for $(q,p,t)\in \mathbb{H}_{1}\mathbf{,}$ $\omega >0$ and $\psi \in L^{2}(%
\mathbb{R})$, called the Schr\"{o}dinger representation. It is square
integrable modulo the center $\mathbb{R}$ of $\mathbb{H}_{1}$ and the Borel
section $\sigma _{0}$ of $\mathbb{H}_{1}$ over $\mathbb{R}^{2}\mathbb{\equiv
H}_{1}/\mathbb{R}$, which is defined by $\sigma _{0}\left( q,p\right)
=\left( q,p,0\right) .$ Therefore, there exists $\left[ 16\right] $ a
self-adjoint, positive semi-invariant operator $\delta $ in $L^{2}\left( 
\mathbb{R}\right) $ such that 
\begin{equation}
\int\limits_{\mathbb{R}^{2}}\left\langle f,T_{\omega }(\sigma _{0}\left(
q,p\right) )\left[ \phi _{1}\right] \right\rangle \left\langle T_{\omega
}(\sigma _{0}\left( q,p\right) )\left[ \phi _{2}\right] ,g\right\rangle
dqdp=\left\langle f,g\right\rangle \left\langle \delta ^{\frac{1}{2}}\phi
_{1},\delta ^{\frac{1}{2}}\phi _{2}\right\rangle ,  \tag{4.40}
\end{equation}%
for all $f,g\in L^{2}(\mathbb{R},d\xi )$\textit{\ and }$\phi _{1},\phi _{2}$
in the domain of $\delta ^{\frac{1}{2}}.$ The group $\mathbb{H}_{1}$ here
being unimodular, therefore the Duflo-Moore opeartor $\delta $ must be a
multiple of the identity operator$.$

\bigskip

Now, one can construct Perelomov's CS\ $\left[ 16\right] $ as orbits of the
unitary operator $T_{\omega }$ acting on a reference state in $L^{2}(\mathbb{%
R)}$. For our purpose, we choose as a fudicial vector an $m$th Hermite
function 
\begin{equation}
\psi _{m}\left( \xi \right) =\left( -1\right) ^{m}\left( \frac{\omega }{\pi }%
\right) ^{\frac{1}{4}}\left( 2^{m}m!\right) ^{-\frac{1}{2}}e^{-\frac{1}{2}%
\omega \xi ^{2}}H_{m}\left( \sqrt{\omega }\xi \right) ,\text{ \ \ }\xi \in 
\mathbb{R},  \tag{4.41}
\end{equation}%
and define the CS by the ket vectors labeled by $\left( x,y\right) \in 
\mathbb{R}^{2}$ as follows%
\begin{equation}
\left\vert \left( x,y\right) ;\omega ,m\right\rangle :=T_{\omega }(\sigma
_{0}(x,\omega y))\left[ \psi _{m}\right] .  \tag{4.42}
\end{equation}%
The latter ones are completely justified by the square integrability
property of $T_{\omega }$ modulo the subgroup $\mathbb{R}$ and the section $%
\sigma _{0}.$ Indeed, we may choose $\phi _{1}=\phi _{2}=\psi _{m}$ in $%
\left( 4.40\right) $ which reduces to%
\begin{equation}
\omega \int\limits_{\mathbb{R}^{2}}\left\langle f,T_{\omega }(\sigma
_{0}\left( x,\omega y\right) )\left[ \psi _{m}\right] \right\rangle
\left\langle T_{\omega }(\sigma _{0}\left( x,\omega y\right) )\left[ \psi
_{m}\right] ,g\right\rangle dxdy=\left\langle f,g\right\rangle \left\langle
\psi _{m},\psi _{m}\right\rangle .  \tag{4.43}
\end{equation}%
Since $\left\langle \psi _{m},\psi _{m}\right\rangle =1$ using the
definition of the CS $\left( 4.42\right) $, Eq.$\left( 4.43\right) $ can be
re-written as%
\begin{equation}
\omega \int\limits_{\mathbb{R}^{2}}\langle f\left\vert \left( x,y\right)
;\omega ,m\right\rangle \langle \left( x,y\right) ;\omega ,m\left\vert
g\right\rangle dxdy=\left\langle f\right\vert \mathbf{1}_{L^{2}(\mathbb{R}%
)}\left\vert g\right\rangle ,  \tag{4.44}
\end{equation}%
for all $f,g\in L^{2}(\mathbb{R},d\xi )$, leading to the resolution of the
identity operator 
\begin{equation}
\mathbf{1}_{L^{2}(\mathbb{R})}=\omega \int\limits_{\mathbb{R}%
^{2}}dxdy\left\vert \left( x,y\right) ;\omega ,m\right\rangle \left\langle
\left( x,y\right) ;\omega ,m\right\vert  \tag{4.45}
\end{equation}%
which should be understood in the weak sense. Setting $z=x+iy\equiv \left(
x,y\right) $ and writing $-x^{2}+ixy$ as $-\frac{1}{2}\left( z^{\ast
2}+zz^{\ast }\right) $, the wave function of the CS in $\left( 4.43\right) $
read 
\begin{equation}
\langle \xi \left\vert z;\omega ,m\right\rangle =\left( -1\right) ^{m}\left( 
\frac{\omega }{\pi }\right) ^{\frac{1}{4}}\left( 2^{m}m!\right) ^{-\frac{1}{2%
}}\exp \left( -\frac{1}{2\omega }z^{\ast 2}+\sqrt{2}\xi z^{\ast }-\frac{1}{%
2\omega }zz^{\ast }-\frac{1}{2}\omega \xi ^{2}\right) H_{m}\left( \sqrt{%
\omega }\xi -\sqrt{\frac{2}{\omega }}\func{Re}z\right) .  \tag{4.46}
\end{equation}%
Its clear that we here recover the previous basis element constructed by the 
$J$-matrix method. That is, 
\begin{equation}
\langle \xi \left\vert z,\omega ,m\right\rangle =\Phi _{m}^{z,\omega }(\xi ).
\tag{4.47}
\end{equation}%
In particular, for $\omega =1$ and $m=0,$ the expression $\left( 4.45\right) 
$ reduce to the CS of the harmonic oscillator given in $\left( 2.2\right) $ 
\begin{equation}
\langle \xi \left\vert z;1,0\right\rangle =\pi ^{-\frac{1}{4}}\exp \left( -%
\frac{1}{2}z^{\ast 2}+\sqrt{2}\xi z^{\ast }-\frac{1}{2}zz^{\ast }-\frac{1}{2}%
\omega \xi ^{2}\right) ,\text{ \ \ }\xi \in \mathbb{R}.  \tag{4.48}
\end{equation}

\subsection{An application to planar Landau levels}

In $\left[ 2\right] $ the authors have been comparing properties of an
operator $\mathfrak{A}$ with symbol $\mathfrak{A}=a\left( \xi ,-i\hbar
\partial _{\xi }\right) $ with those of the operator $\widetilde{\mathfrak{A}%
}=a\left( X^{\gamma ,\mu },Y^{\gamma ,\mu }\right) $ obtained by replacing
formally $\xi $ and $i\hbar \partial _{\xi }$ by the vector fields $%
X^{\gamma ,\mu }=\frac{\gamma }{2}\xi +i\frac{\hbar }{\mu }\partial _{\zeta
} $ and $Y^{\gamma ,\mu }=\frac{\mu }{2}\zeta +i\frac{\hbar }{\gamma }%
\partial _{\xi }$ where $\gamma ,\mu \in \mathbb{R}$ with $\gamma \mu \neq 0.
$ The operator $\mathfrak{A}$ is assumed to be less complicated than its
counterpart $\widetilde{\mathfrak{A}}$ so that one would like to deduce some
properties of the second from the first. As an example when $\mathfrak{A}$
is the harmonic oscillator Hamiltonian%
\begin{equation}
\mathfrak{A=}-\frac{1}{2}\frac{d^{2}}{d\xi ^{2}}+\frac{1}{2}\omega ^{2}\xi
^{2}=H_{har}+\frac{1}{2}\omega ,  \tag{4.49}
\end{equation}%
with $\gamma =1$ and $\mu =\omega ,$ one obtains the operator $\widetilde{%
\mathfrak{A}}$ $=a\left( \frac{1}{2}\xi +i\frac{1}{\omega }\partial _{\zeta
},\frac{\omega }{2}\zeta +i\partial _{\xi }\right) .$ Explicitly,%
\begin{equation}
\widetilde{\mathfrak{A}}=H_{\omega }^{sym}=-\frac{1}{2}\left( \partial _{\xi
}^{2}+\partial _{\zeta }^{2}\right) -i\frac{\omega }{2}\left( \zeta \partial
_{\xi }-\xi \partial _{\zeta }\right) +\frac{1}{2}\left( \frac{\omega }{2}%
\right) ^{2}\left( \xi ^{2}+\zeta ^{2}\right)  \tag{4.50}
\end{equation}%
which is the Landau Hamiltonian describing a charged particle moving in the $%
\left( \xi ,\zeta \right) $ plane under the action of a constant uniforme
magnetic field normal to the plane. Indeed, the authors $\left[ 2\right] $
used a wavepacket transform together with the cross-Wigner transform of the
rescaled Hermite functions $\left( 4.41\right) $ $\left( \omega =1\right) $
to obtain the eigenfunctions in $L^{2}\left( \mathbb{R}^{2}\right) $ of the
two-dimensional magnetic operator $H_{sym}$ associated to the $m$th Landau
level as 
\begin{equation}
\phi _{j+m,m}\left( z\right) =\left( -1\right) ^{j}\frac{1}{\sqrt{2\pi }}%
\left( \frac{j!}{\left( j+m\right) !}\right) ^{\frac{1}{2}}2^{-\frac{1}{2}%
m}z^{m}L_{j}^{\left( m\right) }\left( \frac{1}{2}zz^{\ast }\right) e^{-\frac{%
1}{4}zz^{\ast }}.  \tag{4.51}
\end{equation}%
By intertwining unitarly the operator $\left( 4.49\right) $ via the ground
state transformation as 
\begin{equation}
\Delta _{\omega }:=e^{\frac{\omega }{2}z\overline{z}}\left( \frac{1}{2}%
H_{2\omega }^{sym}-\frac{\omega }{2}\right) e^{-\frac{\omega }{2}z\overline{z%
}}=-\frac{\partial ^{2}}{\partial z\partial \overline{z}}+\omega \overline{z}%
\frac{\partial }{\partial \overline{z}}.  \tag{4.52}
\end{equation}%
The eigenspace of $\Delta _{\omega }$ in $L^{2}\left( \mathbb{C}\text{, }%
e^{-\omega z\overline{z}}d\nu \right) ,$ associated with the $m$th Landau
level 
\begin{equation}
\mathcal{A}_{\omega ,m}\left( \mathbb{C}\right) :=\left\{ F\in L^{2}\left( 
\mathbb{C}\text{, }e^{-\omega z\overline{z}}d\nu \right) ,\Delta _{\omega
}F=\omega mF\right\} .  \tag{4.53}
\end{equation}

Our procedure based on the trigonalization of the harmonic oscillator
together with its shape invariance propertie have provided an orthonormal
basis for this eigenspace as given by the family of vectors $\left(
C_{s}^{\left( m,\omega \right) }\left( z,z^{\ast }\right) \right) _{s\geq
0}. $ For $\omega =1,$ the reproducing kernel of the space $\mathcal{A}%
_{1,m}\left( \mathbb{C}\right) $ which turns out to be the space of $m$-%
\textit{true-polyanalytic} functions $\left[ 18\right] $ may be obtained by
straightforward calculations $\left[ 19\right] $ using Zaremba formula $%
\left[ 20\right] $ as%
\begin{equation}
\sum\limits_{s=0}^{\infty }C_{s}^{\left( m,1\right) }\left( z,z^{\ast
}\right) \left( C_{s}^{\left( m,1\right) }\left( w,w^{\ast }\right) \right)
^{\ast }=\pi ^{-1}e^{zw^{\ast }}L_{m}^{\left( 0\right) }\left( \left\vert
z-w\right\vert ^{2}\right)  \tag{4.54}
\end{equation}%
which is the kernel function of the projection from $L^{2}\left( \mathbb{C}%
\text{, }e^{-z\overline{z}}d\nu \right) $ onto the $m$th Landau eigenspace.

\section{Extension to the Morse oscillator}

\subsection{Review of a known tridiagonalization}

We start by recalling the explicite form of the polynomials $p_{n}\left(
E\right) $ and the density $\Omega \left( E\right) $ associated with the
Morse potential. Its Hamiltonian reads 
\begin{equation}
H_{mor}=-\frac{1}{2}\frac{d^{2}}{dx^{2}}+V_{0}(e^{-2\beta x}-2e^{-\beta x})+%
\frac{1}{2}\beta ^{2}D^{2}  \tag{5.1}
\end{equation}%
and it is acting on the Hilbert space $L^{2}\left( \mathbb{R}\right) $,
where $V_{0}$ and $\beta $ are nonnegative given parameters and $D=\frac{%
\sqrt{2V_{0}}}{\beta }-\frac{1}{2}$. The energy spectrum consists on: $%
\left( i\right) $ a continuous spectrum $\varepsilon \geq \frac{\beta ^{2}}{2%
}D^{2},\left( ii\right) $ a finite discrete set of eigenvalues given by $%
\varepsilon _{\mu }=\frac{\beta ^{2}}{2}\mu (2D-\mu ),$ $\mu =0,1,\ldots
,\lfloor D\rfloor ,$ whose corresponding eigenfunctions are given by 
\begin{equation}
\text{ \ \ }\psi _{\mu }(x)=\sqrt{\frac{\beta (2D-2\mu )\mu !}{\Gamma
(2D-\mu +1)}}y^{D-\mu }e^{-y/2}L_{\mu }^{(2D-2\mu -1)}(y),\text{ }y=\frac{%
\sqrt{2V_{0}}}{\beta }e^{-\beta x},  \tag{5.2}
\end{equation}%
in terms of the Laguerre polynomials $L_{\mu }^{\left( \alpha \right) }$.
Here $\lfloor a\rfloor $ denotes the greatest integer not exceeding $a.$ The
operator $H_{mor}$ has a tridiagonal matrix representation in the complete
orthonormal basis of $L^{2}\left( \mathbb{R}_{+}\right) $%
\begin{equation}
g_{n}(x)=\sqrt{\frac{n!\,\beta }{\Gamma (n+2\gamma +1)}}\,y^{\gamma +\frac{1%
}{2}}\,e^{-\frac{y}{2}}\,L_{n}^{(2\gamma )}(y),\text{ \ }n=0,1,2,...., 
\tag{5.3}
\end{equation}%
where $\gamma $ is a free parameter$.$ The coefficients $\{c_{n},d_{n}%
\}_{n=0}^{\infty }$ which are associated with $H_{mor}$ are given explicitly
as%
\begin{equation}
c_{n}(\gamma )=\frac{\beta }{\sqrt{2}}(n+\gamma +\frac{1}{2}%
-D)\;,\;d_{n}(\gamma )=\frac{-\beta }{\sqrt{2}}\sqrt{n(n+2\gamma )}. 
\tag{5.4}
\end{equation}%
and the polynomials $p_{n}\left( E\right) $ satisfying $\left( 2.5\right) $
were obtained $\left[ 21\right] $ \ as%
\begin{equation}
P_{n}(E)=\frac{(-1)^{n}(\gamma +\frac{1}{2}-D)_{n}}{\sqrt{n!(2\gamma +1)_{n}}%
}{}_{3}F_{2}\left( 
\begin{matrix}
-n,1-D+i\lambda ,1-D+-i\lambda \\ 
\gamma +\frac{1}{2}-D,\gamma +\frac{1}{2}-D%
\end{matrix}%
\mid 1\right)  \tag{5.5}
\end{equation}%
with the orthogonality relations

\begin{equation}
\sum_{\mu =0}^{\lfloor D\rfloor }\Omega _{\mu }\mathcal{P}_{j}(E_{\mu })%
\overline{\mathcal{P}_{k}(E_{\mu })}+\int_{\frac{\alpha ^{2}}{2}%
D^{2}}^{+\infty }\Omega (E)\mathcal{P}_{j}(E)\overline{\mathcal{P}_{k}(E)}%
dE=\delta _{j,k},  \tag{5.6}
\end{equation}%
where $E_{\mu }=2^{-1}\beta ^{2}(\mu (2D-\mu ))$ and the density functions%
\begin{equation}
\Omega _{\mu }=\frac{\Gamma ^{2}(\gamma +\frac{1}{2}+D)}{\Gamma (2D)\Gamma
(2\gamma +1)}\frac{(-2D)_{\mu }(-D+1)_{\mu }((-D+\gamma +\frac{1}{2})_{\mu
})^{2}(-1)^{\mu }}{(-D)_{\mu }((-D-\gamma +\frac{1}{2})_{\mu })^{2}\mu !}, 
\tag{5.7}
\end{equation}%
and 
\begin{equation}
\Omega (E)=\frac{(\alpha ^{2}\lambda _{E})^{-1}}{\Gamma ^{2}(\gamma +\frac{1%
}{2}-D)\Gamma (2\gamma +1)}\left\vert \frac{\Gamma (-D+i\lambda _{E})\Gamma
^{2}(\gamma +\frac{1}{2}+i\lambda _{E})}{2\pi \Gamma (2i\lambda _{E})}%
\right\vert ,  \tag{5.8}
\end{equation}%
with $\ \lambda _{E}^{2}=\frac{1}{2\beta ^{2}}E-D^{2}.$

\subsection{ Glauber-type coherent states for the Morse oscillator $(\protect%
\phi _{0}^{z,\protect\beta ,D})_{z\in \mathbb{C}}$}

In this section, we summarize the construction of the CS we have introduced
in $\left[ 22\right] .$ For that we start by writing the coherent state~$|z)$
of Glauber-type as the normalized solution to the eigenvalue equation 
\begin{equation}
A|z)=z|z).  \tag{5.9}
\end{equation}%
Next, we expand the state $|z)$ in terms of the basis elements $\left\{
g_{n}\right\} $ in $\left( 5.3\right) $ as 
\begin{equation}
|z)=\sum_{n=0}^{\infty }\Lambda _{n}(z)\,{\left\vert g_{n}\right\rangle }. 
\tag{5.10}
\end{equation}%
Inserting the form $\left( 5.10\right) $ into $\left( 5.9\right) $ and using 
$\left( 2.7\right) $, we find that 
\begin{equation}
\sum_{n=0}^{\infty }z\Lambda _{n}(z)\,{\left\vert g_{n}\right\rangle }%
=\sum_{n=0}^{\infty }\left[ c_{n}\Lambda _{n}(z)+\,d_{n+1}\Lambda _{n+1}(z)%
\right] |g_{n}\rangle ,  \tag{5.11}
\end{equation}%
where the coefficients $\{\Lambda _{n}(z)\}_{n=0}^{\infty }$ are given by 
\begin{equation}
\Lambda _{0}(z)=\left( \sum_{n=0}^{\infty }\left\vert Q_{n}(z)\right\vert
^{2}\right) ^{-\frac{1}{2}}\text{and }\Lambda _{n}(z)=\Lambda
_{0}(z)\,Q_{n}(z),\text{ }n\geq 1,  \tag{5.12}
\end{equation}%
with%
\begin{equation}
Q_{0}(z)=1\text{ and}\;Q_{n}(z)=\prod_{j=0}^{n-1}(\frac{z-c_{j}}{d_{j+1}}%
),\;n\geq 1.\;  \tag{5.13}
\end{equation}%
Precisely,%
\begin{equation}
Q_{n}(z)=\prod_{j=0}^{n-1}(\frac{z-c_{j}}{d_{j+1}})=\prod_{j=0}^{n-1}\left( 
\frac{j-\left( \frac{\sqrt{2}}{\beta }z-\left( \gamma +\frac{1}{2}-D\right)
\right) }{\sqrt{\left( j+1\right) \left( j+2\gamma +1\right) }}\right) . 
\tag{5.14}
\end{equation}%
The normalization constant reads%
\begin{equation}
\frac{1}{\left( \Lambda _{0}\left( z\right) \right) ^{2}}=\sum%
\limits_{n=0}^{+\infty }Q_{n}(z)\overline{Q_{n}(z)}=\frac{\Gamma \left(
2\gamma +1\right) \Gamma \left( \frac{2\sqrt{2}}{\beta }\func{Re}z+2D\right) 
}{\left\vert \Gamma \left( \frac{\sqrt{2}}{\beta }z+\gamma +\frac{1}{2}%
+D\right) \right\vert ^{2}}.  \tag{5.15}
\end{equation}%
Returning to $\left( 4.2\right) $ and replacing $Q_{n}\left( z\right) $ and $%
{\left\vert g_{n}\right\rangle }$ by their expressions respectively, we
obtain 
\begin{equation}
\langle x|z)=\Lambda _{0}\left( z\right) \sum_{n=0}^{\infty }Q_{n}(z)\,{%
\left\vert g_{n}\right\rangle }  \tag{5.16}
\end{equation}%
\begin{equation}
=\Lambda _{0}\left( z\right) \sum_{n=0}^{\infty }\left( -\left( \frac{\sqrt{2%
}}{\beta }z-\left( \gamma +\frac{1}{2}-D\right) \right) \right) _{n}\frac{%
\sqrt{\beta \Gamma \left( 2\gamma +1\right) }}{\Gamma (n+2\gamma +1)}%
y^{\gamma +\frac{1}{2}}\,e^{-\frac{y}{2}}\,L_{n}^{(2\gamma )}(y),  \tag{5.17}
\end{equation}%
where $y=\frac{\sqrt{8V_{0}}}{\beta }\,e^{-\beta x}$. A closed form of $%
\left( 4.8\right) $ was obtained $\left[ 22\right] $ as

\begin{equation}
\langle x|z)=C_{z}^{\beta ,\gamma ,D}y^{\frac{\sqrt{2}}{\beta }z+D}e^{-\frac{%
y}{2}}  \tag{5.18}
\end{equation}%
with 
\begin{equation}
C_{z}^{\beta ,\gamma ,D}:=\frac{\sqrt{\beta }}{\sqrt{\Gamma \left( \frac{2%
\sqrt{2}}{\beta }\func{Re}z+2D\right) }}\,\frac{\left\vert \Gamma \left( 
\frac{\sqrt{2}}{\beta }z+\gamma +\frac{1}{2}+D\right) \right\vert }{\Gamma
\left( \gamma +\frac{\sqrt{2}}{\beta }z+\frac{1}{2}+D\right) }  \tag{5.19}
\end{equation}%
provided that $\func{Re}z+\frac{\beta }{\sqrt{2}}D>0.$

\smallskip \smallskip

\textbf{Remark 5.1. }We may compare the wave function $\left( 5.18\right) $
in the $y$ variable with the Klauder wavelet written in the Flandrin's
notation ($\left[ 23\right] $, p.4030$):$

\begin{equation}
\chi \left( f\right) =Cf^{\alpha +i\tau }e^{-\sigma f}U\left( f\right) , 
\tag{5.18}
\end{equation}%
where $C\in \mathbb{C}$ is a constant , $\alpha >0,\sigma >0,\tau \in 
\mathbb{R}$, $U\left( f\right) $ is the unit step function. Indeed, we may
proceed by the following identifications: $C_{z}^{\beta ,\gamma ,D}\equiv C,$
$\chi \left( f\right) \equiv \langle x|z),$ $f\equiv y>0$ is regarded as a 
\textit{frequency }parameter, $\alpha \equiv \frac{\sqrt{2}}{\beta }\func{Re}%
z+D>0$ is satisfied, $\tau \equiv \frac{\sqrt{2}}{\beta }\func{Im}z\in 
\mathbb{R}$ and $\sigma =%
{\frac12}%
>0.$ Following $\left[ 23\right] $, we introduce the function $s\mapsto 
\underline{\chi }\left( s\right) $ where the variable $s$ is the\textbf{\ }%
\textit{scale} in a such way that\textit{\ }$\chi \left( y\right)
=\int\limits_{\mathbb{R}}\underline{\chi }\left( s\right) e^{-2i\pi ys}ds$%
\textit{. }This lead to the following fluctuations%
\begin{equation}
\Delta _{s_{\chi }}^{2}=\frac{1}{E_{\chi }}\int\limits_{\mathbb{R}}\left(
s-s_{0}\right) ^{2}\left\vert \underline{\chi }\left( s\right) \right\vert
^{2}ds,  \tag{5.19}
\end{equation}

\begin{equation}
\Delta _{y_{\chi }}^{2}=\frac{1}{E_{\chi }}\int\limits_{\mathbb{R}}\left(
y-y_{0}\right) ^{2}\left\vert \chi \left( y\right) \right\vert
^{2}y^{2r+1}dy,  \tag{5.20}
\end{equation}%
where 
\begin{equation}
E_{\chi }=\int\limits_{0}^{+\infty }\left\vert \chi \left( y\right)
\right\vert ^{2}y^{2r+1}dy,\text{ \ }s_{0}=\frac{1}{E_{\chi }}\int\limits_{%
\mathbb{R}}s\left\vert \underline{\chi }\left( s\right) \right\vert ^{2}ds,%
\text{ }y_{0}=\frac{1}{E_{\chi }}\int\limits_{\mathbb{R}}y\left\vert \chi
\left( y\right) \right\vert ^{2}dy,  \tag{5.21}
\end{equation}%
and $r$ $\in \mathbb{R}$ is a free parametrer. \ The fluctuations in $\left(
5.19\right) -\left( 5.20\right) $ satisfy the uncertainty relation 
\begin{equation}
\Delta _{s_{\chi }}\Delta _{y_{\chi }}\geq \frac{y_{0}}{4\pi }  \tag{5.22}
\end{equation}%
with equality if and only if $\chi \left( y\right) $ is of form $\left(
4.12\right) $. It is not surprising that the CS $\left( 5.18\right) $ for
the Morse oscillator also minimize an uncertainty relation which is a known
property one may expect CS to satisfy.

\section{Tridiagonalization of the Morse oscillator via the basis $\left( 
\protect\phi _{m}^{z,\protect\beta ,D}\right) _{m\geq 0}$}

In this section, we construct a set of othonormalized functions $\phi
_{m}^{z,\beta ,D},$ $m=0,1,2,...,$ that tridiagonalize the Morse oscillator $%
H_{mor}$. Next we show that these functions may be viewed as generalized CS
in the sense of the formalism in Subsection 2.2. For this, we introduce the
variable 
\begin{equation}
\xi _{z}=D+\frac{\sqrt{2}}{\beta }z-\frac{1}{2}  \tag{6.1}
\end{equation}%
and we denote the above CS in $\left( 5.18\right) $ 
\begin{equation}
\langle x|z)=C_{z}^{\beta ,\gamma ,D}y^{\frac{\sqrt{2}}{\beta }z+D}e^{-\frac{%
1}{2}y}\equiv \phi _{0}^{z,\beta ,D}\left( x\right) =\sqrt{\frac{\beta }{%
\Gamma \left( 2\func{Re}\xi _{z}+1\right) }}y^{\frac{\sqrt{2}}{\beta }%
z+D}e^{-\frac{y}{2}},\ \ \ y=\frac{\sqrt{8V_{0}}}{\beta }\,e^{-\beta x}. 
\tag{6.2}
\end{equation}%
Note that $\phi _{0}^{z,\beta ,D}\left( x\right) $ may also be written as 
\begin{equation}
\phi _{0}^{z,\beta ,D}\left( x\right) =\left( -1\right) ^{0}\sqrt{\frac{%
\beta }{\Gamma \left( 2\func{Re}\xi _{z}+2\right) }}y^{\xi +\frac{1}{2}}e^{-%
\frac{y}{2}}L_{0}^{2\func{Re}\xi _{z}}\left( y\right)  \tag{6.3}
\end{equation}%
in terms of the Laguerre polynomial $L_{0}^{\left( \alpha \right) }\left(
y\right) .$ Starting from $\phi _{0}^{z,\beta ,D},$ our goal is to construct
a set of othonormalized functions $\phi _{m}^{z,\beta ,D},$ $m=0,1,2,...,$%
\begin{equation}
\left\langle \phi _{m}^{z,\beta ,D},\phi _{n}^{z,\beta ,D}\right\rangle
=\delta _{m,n}  \tag{6.4}
\end{equation}%
that tridiagonalize the Morse oscillator $H_{mor},$ and such that the first
vector is the coherent state $\phi _{0}^{z,\beta ,D}\left( x\right) \equiv
\langle x|z).$ For that, we recall the relations 
\begin{equation}
A=-\frac{1}{\sqrt{2}}\frac{d}{dx}+W\left( x\right) ,\text{ \ }A^{\dag }=%
\frac{1}{\sqrt{2}}\frac{d}{dx}+W\left( x\right) ,\text{ \ }  \tag{6.5}
\end{equation}%
where 
\begin{equation}
W\left( x\right) =\frac{\beta }{2\sqrt{2}}\left( y-2D\right) .  \tag{6.6}
\end{equation}%
From $A\phi _{0}^{z,\beta ,D}=z\phi _{0}^{z,\beta ,D}$ and $A^{\dag }\phi
_{0}^{z,\beta ,D}=z^{\ast }\phi _{0}^{z,\beta ,D}+d_{1}\phi _{1}^{z,\beta
,D},$ we deduce that 
\begin{equation}
d_{1}\phi _{1}^{z,\beta ,D}=\left( 2W-\left( z+z^{\ast }\right) \right) \phi
_{0}^{z,\beta ,D}.  \tag{6.7}
\end{equation}%
Taking the square modulus of each term of the equality $\left( 6.7\right) $
and using the orthogonality relations $\left( 6.4\right) $ we obtain that%
\begin{equation}
d_{1}=2\left( \int\limits_{-\infty }^{+\infty }\left[ W\left( x\right) -%
\func{Re}z\right] ^{2}\left\vert \phi _{0}\left( x\right) \right\vert
^{2}dx\right) ^{\frac{1}{2}}=\beta \sqrt{\func{Re}\xi _{z}+\frac{1}{2}}.%
\text{ \ \ }  \tag{6.8}
\end{equation}%
Now, by $\left( 6.7\right) -\left( 6.8\right) ,$ we may extract the second
vector $\phi _{1}^{z,\beta ,D}$ as 
\begin{equation}
\phi _{1}^{z,\beta ,D}\left( x\right) =-\left[ \left( 2\func{Re}\xi
_{z}+1\right) -y\right] \sqrt{\frac{\beta }{\Gamma \left( 2\func{Re}\xi
_{z}+2\right) }}y^{\xi _{z}+\frac{1}{2}}e^{-\frac{y}{2}}  \tag{6.9}
\end{equation}%
which may also be written as 
\begin{equation}
\phi _{1}^{z,\beta ,D}\left( x\right) =\left( -1\right) ^{1}\sqrt{\frac{%
\beta }{\Gamma \left( 2\func{Re}\xi _{z}+2\right) }}y^{\xi _{z}+\frac{1}{2}%
}e^{-\frac{y}{2}}L_{1}^{2\func{Re}\xi _{z}}\left( y\right)  \tag{6.10}
\end{equation}%
in terms of the Laguerre polynomial $L_{1}^{\left( \alpha \right) }\left(
y\right) .$ Once we have determined $\phi _{1}^{z,\beta ,D}$ we can again
proceed by using $\left( 3.7\right) $ to extract the next vector $\phi
_{2}^{z,\beta ,D}$ and so on. More specifically, we may use the principle of
induction to establish (see Appendix B) that the coefficients $\left(
c_{n},d_{n}\right) $ are given by%
\begin{equation}
c_{n}=\frac{\beta }{\sqrt{2}}\left( n+\frac{\sqrt{2}}{\beta }z\right) ,\text{
\ \ }d_{n}=\frac{\beta }{\sqrt{2}}\sqrt{n\left( n+2\func{Re}\xi _{z}\right) }
\tag{6.11}
\end{equation}%
and the vectors basis we are looking for are of the form\textit{\ }%
\begin{equation}
\left\langle x\right\vert z,\beta ,D,m)\equiv \phi _{m}^{z,\beta ,D}\left(
x\right) =\left( -1\right) ^{m}\sqrt{\frac{\beta m!}{\Gamma \left( m+2\func{%
Re}\xi _{z}+1\right) }}y^{\xi _{z}+\frac{1}{2}}e^{-\frac{1}{2}y}L_{m}^{2%
\func{Re}\xi _{z}}\left( y\right) .  \tag{6.12}
\end{equation}%
They form an orthonormal system in the Hilbert space $L^{2}\left( \mathbb{R}%
_{+},y^{-1}dy\right) .$ The states whose wavefunctions are given by $\left(
6.12\right) $ will be called generalized CS of the Morse oscillator with
respect to the index $m=0,1,2,...$ . As we will see their harmonic limit
gives the generalized CS $\Phi _{m}^{z,\omega }$ in $\left( 4.18\right) .$

\smallskip

The family of generalized CS $\left\{ \phi _{n}^{z,\beta ,D}\right\}
_{n=0}^{\infty }$ tridiagonalizes the Hamiltonian $H_{mor}$ as \textit{\ } 
\begin{equation}
H_{mor}\phi _{n}^{z,\beta ,D}=b_{n-1}\phi _{n-1}^{z,\beta ,D}+a_{n}\phi
_{n}^{z,\beta ,D}+b_{n}^{\ast }\phi _{n+1}^{z,\beta ,D},  \tag{6.13}
\end{equation}%
with 
\begin{equation}
a_{n}=\frac{\beta ^{2}}{2}\left( n+\frac{\sqrt{2}}{\beta }z\right) \left( n+%
\frac{\sqrt{2}}{\beta }\overline{z}\right) +\frac{\beta ^{2}}{2}n\left( n+2%
\func{Re}\xi _{z}\right)  \tag{6.14}
\end{equation}%
and%
\begin{equation}
b_{n}=\frac{\beta ^{2}}{2}\left( n+\frac{\sqrt{2}}{\beta }z\right) \sqrt{%
\left( n+1\right) \left( n+2\func{Re}\xi _{z}+1\right) }.  \tag{6.15}
\end{equation}

\subsection{$\left( \protect\phi _{m}^{z,\protect\beta ,D}\right) _{z\in 
\mathbb{C}}$ as generalized CS for the Morse oscillator}

We can show (see Appendix C) that for each fixed $m=0,1,2,...,$ the state $%
\phi _{m}^{z,\beta ,D}$\textit{\ }obey the general form of decomposition $%
\left( 3.14\right) $ as 
\begin{equation}
\phi _{m}^{z,\beta ,D}\left( x\right) =\sum_{s=0}^{N_{b}-1}\sqrt{\left\vert
\Omega \left( \varepsilon _{s}\right) \right\vert }P_{m}\left( \varepsilon
_{s}\right) ^{\ast }\psi _{s}\left( x\right) +\int_{E_{\min }}^{\infty
}\left\vert \Omega \left( \varepsilon \right) \right\vert P_{m}(\varepsilon
)^{\ast }\psi \left( \varepsilon ,x\right) d\varepsilon   \tag{6.16}
\end{equation}%
where%
\begin{equation}
P_{m}\left( u\right) =_{3}\digamma _{2}\left( -m,-D+iu,-D-iu,\frac{\sqrt{2}}{%
\beta }z,\frac{\sqrt{2}}{\beta }z^{\ast };1\right)   \tag{6.17}
\end{equation}%
the density functions are obtained as 
\begin{equation}
\Omega \left( \varepsilon \right) =\left\vert \frac{\Gamma \left(
-D+i\varsigma \left( \varepsilon \right) \right) \Gamma ^{2}\left( \func{Re}%
\xi _{z}+\frac{1}{2}+i\varsigma \left( \varepsilon \right) \right) }{2\pi
\Gamma \left( 2i\varsigma \left( \varepsilon \right) \right) }\right\vert 
\frac{1}{\beta ^{2}\varsigma \left( \varepsilon \right) \Gamma ^{2}\left( 
\func{Re}\xi _{z}+\frac{1}{2}-D\right) \Gamma \left( 2\func{Re}\xi
_{z}+1\right) }  \tag{6.18}
\end{equation}%
and%
\begin{equation}
\Omega \left( \varepsilon _{s}\right) =\frac{\left( -1\right) ^{s}}{s!}\frac{%
\Gamma ^{2}\left( \func{Re}\xi _{z}+\frac{1}{2}+D\right) \left( -2D\right)
_{s}\left[ \left( -D+\func{Re}\xi _{z}+\frac{1}{2}\right) _{s}\right] ^{2}}{%
\Gamma \left( 2D\right) \Gamma \left( 2\func{Re}\xi _{z}+1\right) \left(
-D\right) _{s}\left[ \left( -D-\func{Re}\xi _{z}+\frac{1}{2}\right) _{s}%
\right] ^{2}}  \tag{6.19}
\end{equation}%
where $\varsigma ^{2}\left( \varepsilon \right) =\frac{2}{\beta ^{2}}%
\varepsilon -D^{2}.$ This means that they can be considered as generalized
CS for the $H_{mor}.$ Finally, if we use the definition of the harmonic
limit of the Morse system by adopting, as in $\left[ 24\right] ,$ the
following limiting values of the physical parameters

\begin{equation}
\lim_{HO}\equiv \left\vert 
\begin{array}{c}
V_{0}\rightarrow \infty \\ 
K\rightarrow \infty \\ 
\beta \rightarrow 0 \\ 
\frac{V_{0}}{K}\rightarrow \frac{1}{4}\omega \\ 
V_{0}\beta ^{2}\rightarrow \frac{1}{2}\omega ^{2} \\ 
K\beta ^{2}\rightarrow 2\omega%
\end{array}%
\right.  \tag{6.20}
\end{equation}%
then, we can show (see Appendix D) that, up to a phase factor, we have the
following limit 
\begin{equation}
\lim_{HO}\phi _{m}^{z,\beta ,D}=\Phi _{m}^{z,\omega }  \tag{6.21}
\end{equation}

\medskip

\begin{center}
\textbf{Appendix A}
\end{center}

We start from the sum

\begin{equation}
\mathcal{S}=\sum_{s=0}^{\infty }\left[ \frac{1}{\sqrt{s!}}\frac{z^{\ast s}}{%
\sqrt{\omega }^{s}}L_{m}^{\left( s-m\right) }\left( \frac{1}{\omega }%
zz^{\ast }\right) \right] \psi _{s}\left( \xi \right) ,  \tag{A1}
\end{equation}%
where we replace $\psi _{s}\left( \xi \right) $ by its expession $\left(
4.4\right) $. This gives

\begin{equation}
\mathcal{S}=\left( \frac{\omega }{\pi }\right) ^{\frac{1}{4}}e^{-\frac{1}{2}%
\omega \xi ^{2}}\sum_{s=0}^{\infty }\frac{1}{s!}\left( \frac{z^{\ast }}{%
\sqrt{2\omega }}\right) ^{s}L_{m}^{\left( s-m\right) }\left( \frac{1}{\omega 
}zz^{\ast }\right) H_{s}\left( \sqrt{\omega }\xi \right) =\left[ \left( 
\frac{\omega }{\pi }\right) ^{\frac{1}{4}}e^{-\frac{1}{2}\omega \xi ^{2}}%
\right] \tciFourier  \tag{A2}
\end{equation}%
where 
\begin{equation}
\tciFourier =\sum_{s=0}^{\infty }\frac{1}{s!}\left( \frac{-z^{\ast }}{\sqrt{%
2\omega }}\right) ^{s}L_{m}^{\left( s-m\right) }\left( \frac{1}{\omega }%
zz^{\ast }\right) H_{s}\left( \sqrt{\omega }\xi \right) .  \tag{A3}
\end{equation}%
We set $W=\sqrt{\omega }\xi $ and replace $H_{s}\left( \sqrt{\omega }\xi
\right) $ by the integral representation of the Hermite polynomial $\left( %
\left[ 9\right] \text{, p.105}\right) :$ 
\begin{equation}
H_{s}\left( W\right) =\frac{e^{W^{2}}}{\sqrt{\pi }}\int\limits_{\mathbb{R}%
}\left( -2it\right) ^{s}e^{-t^{2}+2itW}dt  \tag{A4}
\end{equation}%
to rewrite $\left( A3\right) $ as%
\begin{equation}
\tciFourier =\frac{e^{W^{2}}}{\sqrt{\pi }}\int\limits_{\mathbb{R}%
}e^{-t^{2}+2itW}\left[ \sum_{s=0}^{\infty }\frac{1}{s!}\left( \frac{-\sqrt{2}%
itz^{\ast }}{\sqrt{\omega }}\right) ^{s}L_{m}^{\left( s-m\right) }\left( 
\frac{1}{\omega }zz^{\ast }\right) \right] dt  \tag{A5}
\end{equation}%
By using the identity ($\left[ 25\right] $, p.142)$:$%
\begin{equation}
\sum_{k=0}^{\infty }\frac{1}{k!}\left( u\theta \right) ^{k}L_{m}^{\left(
k-m\right) }\left( u\right) =\frac{u^{m}}{m!}\left( \theta -1\right)
^{m}e^{u\theta }  \tag{A6}
\end{equation}%
for $u=\frac{1}{\omega }zz^{\ast }$, $\frac{1}{\omega }zz^{\ast }\theta =%
\frac{-\sqrt{2}itz^{\ast }}{\sqrt{\omega }}$ and $\theta =\frac{-it\sqrt{%
2\omega }}{z},$ Eq.$\left( A5\right) $ takes the form

\begin{equation}
\tciFourier =\frac{\left( -1\right) ^{m}}{m!}\left( \frac{1}{\omega }z^{\ast
}\right) ^{m}\frac{e^{W^{2}}}{\sqrt{\pi }}\int\limits_{\mathbb{R}}\left( it%
\sqrt{2\omega }+z\right) ^{m}\exp \left( -t^{2}-\left( -2iW+\frac{i\sqrt{2}%
z^{\ast }}{\sqrt{\omega }}\right) t\right) dt.  \tag{A7}
\end{equation}%
We use the binomial formula%
\begin{equation}
\left( it\sqrt{2\omega }+z\right) ^{m}=\sum\limits_{j=0}^{m}C_{m}^{j}\left(
it\sqrt{2\omega }\right) ^{j}z^{m-j}  \tag{A8}
\end{equation}%
to rewrite the quantity $\tciFourier $ as 
\begin{equation}
\tciFourier =\frac{\left( -1\right) ^{m}}{m!}\left( \frac{1}{\omega }z^{\ast
}\right) ^{m}\frac{e^{W^{2}}}{\sqrt{\pi }}\Re  \tag{A.9}
\end{equation}%
where%
\begin{equation}
\Re =\int\limits_{\mathbb{R}}\left( \sum\limits_{j=0}^{m}C_{m}^{j}\left( it%
\sqrt{2\omega }\right) ^{j}z^{m-j}\right) \exp \left( -t^{2}-\left( -2iW+%
\frac{i\sqrt{2}z^{\ast }}{\sqrt{\omega }}\right) t\right)
dt=z^{m}\sum\limits_{j=0}^{m}C_{m}^{j}\left( \frac{i\sqrt{2\omega }}{z}%
\right) ^{j}\tciLaplace _{j}.  \tag{A10}
\end{equation}%
To compute the integral%
\begin{equation}
\tciLaplace _{j}=\int\limits_{\mathbb{R}}t^{j}\exp \left( -t^{2}-\left( -2iW+%
\frac{i\sqrt{2}z^{\ast }}{\sqrt{\omega }}\right) t\right) dt  \tag{A11}
\end{equation}%
we use the identity ($\left[ 26\right] ,$p.344)$:$%
\begin{equation}
\int\limits_{\mathbb{R}}t^{j}e^{-pt^{2}-qt}dt=\left( \frac{i}{2}\right) ^{j}%
\sqrt{\pi }p^{-\frac{\left( j+1\right) }{2}}\exp \left( \frac{q^{2}}{4p}%
\right) H_{j}\left( \frac{iq}{2\sqrt{p}}\right) .  \tag{A12}
\end{equation}%
This gives%
\begin{equation}
\tciLaplace _{j}=\left( \frac{i}{2}\right) ^{j}\sqrt{\pi }\exp \left( \frac{1%
}{4}\left( -2iW+\frac{i\sqrt{2}z^{\ast }}{\sqrt{\omega }}\right) ^{2}\right)
H_{j}\left( \frac{1}{2}i\left( -2iW+\frac{i\sqrt{2}z^{\ast }}{\sqrt{\omega }}%
\right) \right)  \tag{A13}
\end{equation}%
Therefore, 
\begin{equation}
\Re =z^{m}\sqrt{\pi }\exp \left( \left( \frac{iz^{\ast }}{\sqrt{2\omega }}%
-iW\right) ^{2}\right) \sum\limits_{j=0}^{m}C_{m}^{j}\left( \frac{-\sqrt{%
\omega }}{z\sqrt{2}}\right) ^{j}H_{j}\left( W-\frac{z^{\ast }}{\sqrt{2\omega 
}}\right) =z^{m}\sqrt{\pi }\exp \left( \frac{1}{4}\left( \frac{i\sqrt{2}%
z^{\ast }}{\sqrt{\omega }}-2iW\right) ^{2}\right) \wp  \tag{A14}
\end{equation}%
where 
\begin{equation}
\wp =\sum\limits_{j=0}^{m}C_{m}^{j}\left( \frac{-\sqrt{\omega }}{z\sqrt{2}}%
\right) ^{j}H_{j}\left( W-\frac{z^{\ast }}{\sqrt{2\omega }}\right)  \tag{A15}
\end{equation}%
By using the summation formula ($\left[ 27\right] $, p.640)$:$%
\begin{equation}
\sum\limits_{j=0}^{m}C_{m}^{j}t^{j}H_{j}\left( X\right) =t^{m}H_{m}\left( X+%
\frac{1}{2t}\right) ,  \tag{A16}
\end{equation}%
Eq.$\left( A15\right) $ reduces to%
\begin{equation}
\wp =\left( \frac{-\sqrt{\omega }}{z\sqrt{2}}\right) ^{m}H_{m}\left( \sqrt{%
\omega }\xi -\sqrt{\frac{2}{\omega }}\func{Re}z\right) ,\text{ }W=\sqrt{%
\omega }\xi .  \tag{A17}
\end{equation}%
Summarizing the above calculations we arrive at the following closed form
for the sum $\left( A1\right) :$ 
\begin{equation*}
\mathcal{S}=\left( \frac{\omega }{\pi }\right) ^{\frac{1}{4}}e^{\frac{1}{2}%
\omega \xi ^{2}}\frac{1}{m!}z^{\ast m}\left( \frac{1}{\sqrt{2\omega }}%
\right) ^{m}\exp \left( \left( \frac{iz^{\ast }}{\sqrt{2\omega }}-iW\right)
^{2}\right) H_{m}\left( W-\sqrt{\frac{2}{\omega }}\func{Re}z\right)
\end{equation*}

\medskip

\begin{center}
\textbf{Appendix B}

\medskip
\end{center}

We assume the following forms for the coefficients $\left(
c_{n},d_{n}\right) :$%
\begin{equation}
c_{n}=\frac{\beta }{\sqrt{2}}\left( n+\frac{\sqrt{2}}{\beta }z\right) ,\text{
\ \ }d_{n}=\frac{\beta }{\sqrt{2}}\sqrt{n\left( n+2\func{Re}\xi _{z}\right) }
\tag{B1}
\end{equation}%
and the vectors basis we are looking for are of the form\textit{\ }%
\begin{equation}
\phi _{n}^{z,\beta ,D}\left( x\right) =\left( -1\right) ^{n}\sqrt{\frac{%
\beta n!}{\Gamma \left( n+2\func{Re}\xi _{z}+1\right) }}y^{\xi _{z}+\frac{1}{%
2}}e^{-\frac{1}{2}y}L_{n}^{2\func{Re}\xi _{z}}\left( y\right) ,\text{ }y=%
\frac{\sqrt{8V_{0}}}{\beta }\,e^{-\beta x},  \tag{B2}
\end{equation}%
for all $n\leq k.$ We will show that it is true for the case $n=k+1.$ For
that we recall the relations $A^{\dagger }\phi _{k}=c_{k}^{\ast }\phi
_{k}+d_{k+1}^{\ast }\phi _{k+1}$ and $A\phi _{k}=c_{k}\phi _{k}+d_{k}\phi
_{k}$ which lead to 
\begin{equation}
\left( A^{\dagger }+A\right) \phi _{k}=2\func{Re}\left( c_{k}\phi
_{k}\right) +d_{k}\phi _{k}+d_{k}\phi _{k-1}+d_{k+1}^{\ast }\phi _{k+1}. 
\tag{B3}
\end{equation}%
From $\left( 6.5\right) $-$\left( 6.6\right) ,$ Eq.$\left( B1\right) $ gives%
\begin{equation}
d_{k+1}^{\ast }\phi _{k+1}=2\left( W-\func{Re}c_{k}\right) \phi
_{k}-d_{k}\phi _{k-1}.  \tag{B4}
\end{equation}%
Explicitly,%
\begin{equation}
d_{k+1}^{\ast }\phi _{k+1}\left( x\right) =\left( -1\right) ^{k+1}\sqrt{%
\frac{\beta \left( k+1\right) !}{\Gamma \left( k+2\func{Re}\xi _{z}+2\right) 
}}y^{\xi _{z}+\frac{1}{2}}e^{-\frac{1}{2}y}\frac{\beta }{\sqrt{2}}\sqrt{%
\left( k+1\right) \left( k+2\func{Re}\xi _{z}+1\right) }\digamma \left(
y\right)  \tag{B5}
\end{equation}%
where%
\begin{equation}
\digamma \left( y\right) =-\frac{2\sqrt{2}}{\beta \left( k+1\right) }\left(
W\left( x\right) -\func{Re}c_{k}\right) L_{k}^{2\func{Re}\xi _{z}}\left(
y\right) +\frac{\sqrt{2}}{\beta \left( k+1\right) }\sqrt{\frac{k+2\func{Re}%
\xi _{z}}{k}}d_{k}L_{k-1}^{2\func{Re}\xi _{z}}\left( y\right)  \tag{B6}
\end{equation}%
By making use of the recursion relations for the Laguerre polynomials with
respect to the degree ($\left[ 25\right] $, p.137) :%
\begin{equation}
\left( n+\alpha \right) L_{n-1}^{\left( \alpha \right) }\left( u\right)
+\left( n+1\right) L_{n+1}^{\left( \alpha \right) }\left( u\right) -\left(
\alpha +1-u+2n\right) L_{n}^{\left( \alpha \right) }\left( u\right) =0 
\tag{B7}
\end{equation}%
Eq.$\left( B5\right) $ becomes%
\begin{equation}
d_{k+1}^{\ast }\phi _{k+1}\left( x\right) =\frac{\beta }{\sqrt{2}}\sqrt{%
\left( k+1\right) \left( k+2\func{Re}\xi _{z}+1\right) }G_{k+1}\left(
y\right)  \tag{B8}
\end{equation}%
where%
\begin{equation}
G_{j}\left( y\right) =\left( -1\right) ^{j}\sqrt{\frac{\beta \left(
j+1\right) !}{\Gamma \left( j+2\func{Re}\xi _{z}+1\right) }}y^{\xi _{z}+%
\frac{1}{2}}L_{j}^{2\func{Re}\xi _{z}}\left( y\right)  \tag{B9}
\end{equation}%
From the norms $\left\langle G_{j},G_{j}\right\rangle =1$ and $\left\langle
\phi _{r},\phi _{r}\right\rangle =1,$ we deduce that $d_{k+1}=d_{k+1}^{\ast
}=\frac{\beta }{\sqrt{2}}\sqrt{\left( k+1\right) \left( k+2\func{Re}\xi
_{z}+1\right) }$ and $\phi _{k+1}\left( x\right) =G_{k+1}\left( y\right) $.
We now use the relation $c_{k+1}=\left\langle \phi _{k+1}\left\vert
A\right\vert \phi _{k+1}\right\rangle $ together with the above results in
order to compute the following integral 
\begin{equation}
c_{k+1}=\int\limits_{-\infty }^{+\infty }\phi _{k+1}\left( \xi \right)
^{\ast }\left[ \frac{-1}{\sqrt{2}}\frac{d}{d\xi }+W\left( \xi \right) \right]
\phi _{k+1}\left( \xi \right) d\xi  \tag{B10}
\end{equation}%
which gives 
\begin{equation}
c_{k+1}=\frac{\beta }{\sqrt{2}}\left( k+1+\frac{\sqrt{2}}{\beta }z\right) 
\tag{B11}
\end{equation}

\medskip

\begin{center}
\textbf{Appendix C}

\medskip
\end{center}

We start from the decomposition of eigenfunctions of $H$ over the basis $%
\left\{ \phi _{m}^{z,\beta ,D}\right\} _{m=0}^{\infty }$ as follows 
\begin{equation}
\psi \left( \varepsilon ,x\right) =\sqrt{\Omega \left( \varepsilon \right) }%
\sum\limits_{m=0}^{\infty }p_{m}\left( \varepsilon \right) \phi
_{m}^{z,\beta ,D}\left( x\right)  \tag{C1}
\end{equation}%
where the coefficients $p_{m}\left( \varepsilon \right) $ satisfy the three
terms recursion relation%
\begin{equation}
\varepsilon p_{m}\left( \varepsilon \right) =b_{m-1}p_{m-1}\left(
\varepsilon \right) +a_{m}p_{m}\left( \varepsilon \right) +b_{m}^{\ast
}p_{m+1}\left( \varepsilon \right)  \tag{C2}
\end{equation}%
with $p_{0}\left( \varepsilon \right) =1$ and%
\begin{equation}
\sum\limits_{m=0}^{N_{b}-1}\left\vert \Omega \left( \varepsilon _{s}\right)
\right\vert p_{m}\left( \varepsilon _{s}\right) p_{m}^{\ast }\left(
\varepsilon _{s}\right) +\int\limits_{E_{\min }}^{+\infty }\left\vert \Omega
\left( \varepsilon \right) \right\vert p_{m}\left( \varepsilon \right)
p_{m}^{\ast }\left( \varepsilon \right) d\varepsilon =\delta _{n,m}. 
\tag{C3}
\end{equation}%
Here $N_{b}=\left\lfloor D\right\rfloor +1$ is the number of bound states of
the Morse oscillator. Also, $E_{\min }=\frac{1}{2}\beta ^{2}D^{2}$ is the
minimum energy of the continous part of the energy spectrum. We may, then,
rescale the energy $\varepsilon $ in terms of the variable $\varsigma $ such
that $\varepsilon =\frac{1}{2}\beta ^{2}\left( D^{2}+\varsigma ^{2}\right) .$
By setting $p_{m}\left( \varepsilon \right) =p_{m}\left( 0\right)
q_{m}\left( \varepsilon \right) ,$ $q_{m}\left( 0\right) =q_{0}\left(
\varepsilon \right) =1,$\ \ Eq.$\left( C2\right) $ takes the form 
\begin{equation}
\varepsilon p_{m}\left( \varepsilon \right) =b_{m-1}\left( \frac{%
p_{m-1}\left( 0\right) }{p_{m}\left( 0\right) }\right) q_{m-1}\left(
\varepsilon \right) +a_{m}q_{m}\left( \varepsilon \right) +b_{m}^{\ast
}\left( \frac{p_{m+1}\left( 0\right) }{p_{m}\left( 0\right) }\right)
q_{m+1}\left( \varepsilon \right) .  \tag{C4}
\end{equation}%
We know that the coefficients $\left\{ c_{n},d_{n}\right\} _{n=0}^{\infty }$
are related to the polynomials $\left\{ p_{n}\left( 0\right) \right\}
_{n=0}^{\infty }$ by the relation%
\begin{equation}
c_{m}c_{m}^{\ast }=-b_{m}^{\ast }\frac{p_{m+1}\left( 0\right) }{p_{m}\left(
0\right) },\text{ \ }d_{m}d_{m}^{\ast }=-b_{m}\frac{p_{m-1}\left( 0\right) }{%
p_{m}\left( 0\right) }.  \tag{C5}
\end{equation}%
Then $\left( C4\right) $ becomes 
\begin{equation}
\varepsilon p_{m}\left( \varepsilon \right) =-\left\vert d_{m}\right\vert
^{2}q_{m-1}\left( \varepsilon \right) +\left( \left\vert c_{m}\right\vert
^{2}+\left\vert d_{m}\right\vert ^{2}\right) q_{m}\left( \varepsilon \right)
-\left\vert c_{m}\right\vert ^{2}q_{m+1}\left( \varepsilon \right)  \tag{C6}
\end{equation}%
In terms of the variable $\varsigma $%
\begin{equation}
-\left( D^{2}+\varsigma ^{2}\right) q_{m}\left( \varsigma \right) =-m\left(
m+2\func{Re}\xi _{z}\right) q_{m-1}\left( \varsigma \right)  \tag{C7}
\end{equation}%
\begin{equation*}
+\left[ \left( m+\frac{\sqrt{2}}{\beta }z\right) \left( m+\frac{\sqrt{2}}{%
\beta }z^{\ast }\right) +m\left( m+2\func{Re}\xi _{z}\right) \right]
q_{m}\left( \varsigma \right) -\left( m+\frac{\sqrt{2}}{\beta }z\right)
\left( m+\frac{\sqrt{2}}{\beta }z^{\ast }\right) q_{m+1}\left( \varsigma
\right)
\end{equation*}%
This is the recursion equation of the continous dual Hahn polynomial $\left[
10\right] .$ The solution is given by 
\begin{equation}
q_{m}\left( \varsigma \right) =_{3}\digamma _{2}\left( -m,a+i\varsigma
,a-i\varsigma ;a+b,a+c;1\right)  \tag{C8}
\end{equation}%
where $a=-D,b=D+\frac{\sqrt{2}}{\beta }z,$ $c=D+\frac{\sqrt{2}}{\beta }%
z^{\ast }$ and

\begin{equation}
\frac{p_{m+1}\left( 0\right) }{p_{m}\left( 0\right) }=-\frac{\left( m+\frac{%
\sqrt{2}}{\beta }z\right) }{\sqrt{\left( m+1\right) \left( m+\func{Re}\xi
_{z}+1\right) }}  \tag{C9}
\end{equation}%
with%
\begin{equation}
p_{m}\left( 0\right) =\left( -1\right) ^{m}\frac{1}{\sqrt{m!}}\sqrt{\frac{%
\Gamma \left( \func{Re}\xi _{z}+1\right) }{\Gamma \left( m+\func{Re}\xi
_{z}+1\right) }}\frac{\Gamma \left( m+\frac{\sqrt{2}}{\beta }z\right) }{%
\Gamma \left( \frac{\sqrt{2}}{\beta }z\right) }  \tag{C10}
\end{equation}%
Therefore,%
\begin{equation}
p_{m}\left( \varsigma \right) =\frac{\left( -1\right) ^{m}}{\sqrt{m!}}\sqrt{%
\frac{\Gamma \left( \func{Re}\xi _{z}+1\right) }{\Gamma \left( m+\func{Re}%
\xi _{z}+1\right) }}\frac{\Gamma \left( m+\frac{\sqrt{2}}{\beta }z\right) }{%
\Gamma \left( \frac{\sqrt{2}}{\beta }z\right) }._{3}\digamma _{2}\left(
-m,-D+i\varsigma ,-D-i\varsigma ,\frac{\sqrt{2}}{\beta }z,\frac{\sqrt{2}}{%
\beta }z^{\ast };1\right) .  \tag{C11}
\end{equation}%
The density $\Omega \left( \varepsilon \right) $ associated with continuous
part of the energy spectrum and $\Omega \left( \varepsilon _{s}\right) $
associated with the bound states $\left\{ \varepsilon _{s}\right\} $ are
found explicitly in $\left[ 21\right] $ as 
\begin{equation}
\Omega \left( \varepsilon \right) =\left\vert \frac{\Gamma \left(
-D+i\varsigma \left( \varepsilon \right) \right) \Gamma ^{2}\left( \func{Re}%
\xi _{z}+\frac{1}{2}+i\varsigma \left( \varepsilon \right) \right) }{2\pi
\Gamma \left( 2i\varsigma \left( \varepsilon \right) \right) }\right\vert 
\frac{1}{\beta ^{2}\varsigma \left( \varepsilon \right) \Gamma ^{2}\left( 
\func{Re}\xi _{z}+\frac{1}{2}-D\right) \Gamma \left( 2\func{Re}\xi
_{z}+1\right) }  \tag{C12}
\end{equation}%
and%
\begin{equation}
\Omega \left( \varepsilon _{s}\right) =\frac{\left( -1\right) ^{s}}{s!}\frac{%
\Gamma ^{2}\left( \func{Re}\xi _{z}+\frac{1}{2}+D\right) \left( -2D\right)
_{s}\left[ \left( -D+\func{Re}\xi _{z}+\frac{1}{2}\right) _{s}\right] ^{2}}{%
\Gamma \left( 2D\right) \Gamma \left( 2\func{Re}\xi _{z}+1\right) \left(
-D\right) _{s}\left[ \left( -D-\func{Re}\xi _{z}+\frac{1}{2}\right) _{s}%
\right] ^{2}}.  \tag{C13}
\end{equation}%
Therefore, we have may expand any eigenfunction $\psi _{s}$ of $H_{mor}$
over the generalized CS basis $\phi _{m}^{z,\beta ,D}\left( x\right) $ as%
\begin{equation}
\psi _{s}\left( x\right) =\sqrt{\Omega \left( \varepsilon _{s}\right) }%
\sum\limits_{m=0}^{\infty }p_{m}\left( \varepsilon _{s}\right) \phi
_{m}^{z,\beta ,D}\left( x\right)  \tag{C14}
\end{equation}%
for bound states, and 
\begin{equation}
\psi \left( \varepsilon ,x\right) =\sqrt{\Omega \left( \varepsilon \right) }%
\sum\limits_{m=0}^{\infty }p_{m}\left( \varepsilon \right) \phi
_{m}^{z,\beta ,D}\left( x\right)  \tag{C15}
\end{equation}%
for scattering states. Thus, we may expand $\phi _{m}^{z,\beta ,D}\left(
x\right) $ over all these eigenstates as%
\begin{equation}
\phi _{m}^{z,\beta ,D}\left( x\right) =\sum\limits_{s=0}^{N_{b}-1}\sqrt{%
\Omega \left( \varepsilon _{s}\right) }p_{m}^{\ast }\left( \varepsilon
_{s}\right) \psi _{s}\left( x\right) +\int\limits_{E_{\min }}^{\infty
}\Omega \left( \varepsilon \right) p_{m}^{\ast }\left( \varepsilon \right)
\psi \left( \varepsilon ,x\right) d\varepsilon .  \tag{C16}
\end{equation}

\medskip

\begin{center}
\textbf{Appendix D}

\medskip
\end{center}

\bigskip We recall the following regarding the Morse Hamiltonian. The
potential is given by 
\begin{equation}
V\left( x\right) =V_{0}\left( e^{-2\beta x}-2e^{-\beta x}\right) +\frac{1}{2}%
\beta ^{2}D^{2}  \tag{D1}
\end{equation}%
where 
\begin{equation}
D=\frac{\sqrt{2V_{0}}}{\beta }-\frac{1}{2}=\frac{1}{2}\left( K-1\right) ,%
\text{ \ }y=\frac{\sqrt{8V_{0}}}{\beta }e^{-\beta x}=Ke^{-\beta x}.  \tag{D2}
\end{equation}%
We use Popov's $\left[ 24\right] $ definition of the harmonic limit of the
Morse system by adopting the following limiting values of the physical
parameters 
\begin{equation}
\lim_{HO}\equiv \left\vert 
\begin{array}{c}
V_{0}\rightarrow \infty \\ 
K\rightarrow \infty \\ 
\beta \rightarrow 0 \\ 
\frac{V_{0}}{K}\rightarrow \frac{1}{4}\omega \\ 
V_{0}\beta ^{2}\rightarrow \frac{1}{2}\omega ^{2} \\ 
K\beta ^{2}\rightarrow 2\omega%
\end{array}%
\right. .  \tag{D3}
\end{equation}%
Recalling Eq.$\left( 6.1\right) ,$ we see that of the variable $\xi _{z}$
satisfies $2\func{Re}\xi _{z}=K+\frac{2\sqrt{2}}{\beta }\func{Re}z-2.$ By
introducing the variable%
\begin{equation}
Y=\frac{y-2\func{Re}\xi }{\sqrt{2\func{Re}\xi }},  \tag{D4}
\end{equation}%
then 
\begin{equation}
Y\equiv \frac{\left[ -K\beta ^{2}x+\frac{1}{2}K\beta ^{3}x^{2}-...\right] %
\left[ 2\sqrt{2}\func{Re}z-2\beta \right] }{\sqrt{K\beta ^{2}+2\beta \sqrt{2}%
\func{Re}z-2\beta ^{2}}}  \tag{D5}
\end{equation}%
One may check that%
\begin{equation}
\lim_{HO}Y\equiv -\sqrt{2\omega }\left( x+\frac{\sqrt{2}}{\omega }\func{Re}%
z\right)  \tag{D6}
\end{equation}%
By another hand, we may apply the following limit for the Laguerre
polynomial $\left( \left[ 28\right] \right) :$ 
\begin{equation}
\left( -1\right) ^{n}n!2^{\frac{1}{2}n}\lim_{\alpha \rightarrow \infty }%
\left[ \alpha ^{-\frac{1}{2}n}L_{n}^{\left( \alpha \right) }\left( \alpha +u%
\sqrt{2\alpha }\right) \right] =H_{n}\left( u\right)  \tag{D7}
\end{equation}%
This gives%
\begin{equation}
\lim_{HO}\left( 2\func{Re}\xi _{z}+1\right) ^{-\frac{1}{2}m}L_{m}^{\left( 
\func{Re}\xi _{z}\right) }\left( y\right) =\frac{1}{m!\sqrt{2^{m}}}%
H_{m}\left( \sqrt{\omega }\left( x+\frac{\sqrt{2}}{\omega }\func{Re}\xi
_{z}\right) \right) .  \tag{D8}
\end{equation}%
Now, we consider the form of the basis that tridiagonalizes the Morse
Hamiltonian 
\begin{equation}
\phi _{m}^{z,\beta ,D}\left( x\right) =\left( -1\right) ^{m}\sqrt{\frac{%
\beta m!}{\Gamma \left( m+2\func{Re}\xi _{z}+1\right) }}y^{\xi _{z}+\frac{1}{%
2}}e^{-\frac{1}{2}y}L_{m}^{2\func{Re}\xi _{z}}\left( y\right) ,\text{ \ }\xi
_{z}=\left( D+\frac{\sqrt{2}}{\beta }z\right) -\frac{1}{2}.  \tag{D9}
\end{equation}%
$\left( a\right) $ We use consecutively, the two asymptotic relations for
very large $\left\vert \xi _{z}\right\vert \rightarrow \infty ,$ namely,%
\begin{equation}
\Gamma \left( m+\func{Re}\xi _{z}+1\right) \sim \left( \func{Re}\xi
_{z}+1\right) ^{m}\Gamma \left( \func{Re}\xi _{z}+1\right)  \tag{D10}
\end{equation}%
\begin{equation}
\Gamma \left( \func{Re}\xi _{z}+1\right) \sim \sqrt{2\pi }\func{Re}\xi _{z}^{%
\func{Re}\xi _{z}-\frac{1}{2}}e^{-\func{Re}\xi _{z}}  \tag{D11}
\end{equation}%
to write%
\begin{equation}
\sqrt{\frac{\beta }{\Gamma \left( m+2\func{Re}\xi _{z}+1\right) }}\sim \sqrt{%
\sqrt{\frac{2\beta ^{2}\func{Re}\xi _{z}}{2\pi }}}\left( 2\func{Re}\xi
_{z}+1\right) ^{-\frac{1}{2}n}\left( 2\func{Re}\xi _{z}\right) ^{-\func{Re}%
\xi _{z}-\frac{1}{2}}e^{\func{Re}\xi _{z}}  \tag{D12}
\end{equation}%
$\left( b\right) $ $\beta ^{2}\left( 2\func{Re}\xi \right) =_{HO}=\beta
^{2}K=2\omega ,$%
\begin{equation}
y^{\xi _{z}+\frac{1}{2}}e^{-\frac{1}{2}y}=K^{\xi _{z}+\frac{1}{2}}e^{-\beta
x\left( \xi _{z}+\frac{1}{2}\right) -\frac{1}{2}Ke^{-\beta x}}  \tag{D13}
\end{equation}%
$\left( c\right) =K^{\xi +\frac{1}{2}}e^{-\beta x}\left( \frac{K}{2}+\frac{%
\sqrt{2}}{\beta }z-1\right) -\frac{1}{2}K\left( 1-\beta x+\frac{1}{2}\beta
^{2}x^{2}+...\right) $

=$_{HO}=e^{\frac{1}{\omega }z^{2}}K^{\xi +\frac{1}{2}}e^{-\frac{1}{2}K}e^{-%
\frac{1}{2}\omega \left( x+\frac{\sqrt{2}}{\omega }z\right) ^{2}}$

\smallskip

To summarize%
\begin{equation}
\phi _{m}^{z,\beta ,D}\left( x\right) =\lim_{HO}\left( -1\right) ^{m}\sqrt{m!%
}\sqrt{\sqrt{\frac{\omega }{\pi }}}e^{\frac{Z^{2}}{\omega }}\left[ e^{-\frac{%
1}{2}K}e^{\func{Re}\xi }\right] \left[ K^{\xi +\frac{1}{2}}\left( 2\func{Re}%
\xi \right) ^{-\func{Re}\xi -\frac{1}{2}}\right] \left( 2\func{Re}\xi
+1\right) ^{-\frac{m}{2}}L_{m}^{2\func{Re}\xi _{z}}\left( y\right) , 
\tag{D14}
\end{equation}%
Furthermore%
\begin{equation}
\lim_{HO}\left[ e^{-\frac{1}{2}K}e^{\func{Re}\xi }\right] =e^{\frac{\sqrt{2}%
}{\beta }\func{Re}z-1}  \tag{D15}
\end{equation}%
\begin{equation}
\lim_{HO}\left[ K^{\xi +\frac{1}{2}}\left( 2\func{Re}\xi \right) ^{-\func{Re}%
\xi -\frac{1}{2}}e^{\func{Re}\xi }\right] =K^{i\frac{\sqrt{2}}{\beta }\func{%
Im}z}\frac{1}{\left( 1+\frac{2\sqrt{2}}{\beta }\func{Re}z-\frac{2}{K}\right)
^{\frac{K}{2}+\frac{\sqrt{2}}{\beta }\func{Re}z-\frac{1}{2}}}  \tag{D16}
\end{equation}%
The first term is just a phase factor. As for the second term, we use the
limiting relation for very small $\delta $ and large $M,$ we can write 
\begin{equation}
Log\left( 1+\delta \right) ^{M}=MLog\left( 1+\delta \right) =M\left( \delta -%
\frac{1}{2}\delta ^{2}+...\right)  \tag{D17}
\end{equation}%
Thus, 
\begin{equation}
\lim_{HO}Log\left[ \left( 1+\frac{2\sqrt{2}}{\beta }\func{Re}z-\frac{2}{K}%
\right) ^{\frac{K}{2}+\frac{\sqrt{2}}{\beta }\func{Re}z-\frac{1}{2}}\right] =%
\frac{\sqrt{2}}{\beta }\func{Re}z-1+\frac{\left( \func{Re}z\right) ^{2}}{%
\omega }  \tag{D18}
\end{equation}%
Therefore,%
\begin{equation}
\lim_{HO}\left( 1+\frac{2\sqrt{2}}{\beta }\func{Re}z-\frac{2}{K}\right) ^{%
\frac{K}{2}+\frac{\sqrt{2}}{\beta }\func{Re}z-\frac{1}{2}}=\exp \left( -%
\frac{\sqrt{2}}{\beta }\func{Re}z+1-\frac{\left( \func{Re}z\right) ^{2}}{%
\omega }\right)  \tag{D19}
\end{equation}%
Thus, combining tha above results together with $\left( 6.8\right) ,$ we
get, up to a phase factor,%
\begin{equation}
\lim_{HO}\phi _{m}^{z,\beta ,D}\left( x\right) \propto \left( -1\right) ^{m}%
\sqrt{\sqrt{\frac{\omega }{\pi }}}\frac{1}{\sqrt{m!2^{m}}}e^{\frac{1}{%
2\omega }\left( z^{2}-\left\vert z\right\vert ^{2}\right) }\exp \left( -%
\frac{1}{2}\omega \left( x+\frac{\sqrt{2}}{\omega }z\right) ^{2}\right)
H_{m}\left( \sqrt{\omega }\left( x+\frac{\sqrt{2}}{\omega }\func{Re}z\right)
\right)  \tag{D20}
\end{equation}%
which may also be written as 
\begin{equation}
\lim_{HO}\phi _{m}^{z,\beta ,D}\left( x\right) \propto \Phi _{m}^{-z^{\ast
},\omega }(x).  \tag{D21}
\end{equation}

\medskip

\begin{center}
\textbf{References}
\end{center}

\begin{quote}
\medskip

$\left[ 1\right] $ S. T. Ali, J. P. Antoine and J. P. Gazeau, Coherent
States, Wavelets and Their Generalizations, second edition, Springer
Science+Business Media New York, 2014

$\left[ 2\right] $ M. de Gosson and F. Luef, Spectral and Regularity
properties of a Pseudo-Differential Calculus Related to Landau Quantization,
J. Pseudo-Diff. Oper. App., 1 (1) (2010) 3-34

$\left[ 3\right] $ Schr\"{o}dinger E, Der stretige Ubergang von der
Mikro-zur Makromechanik, Naturwissenschaften. 14, pp.664-666 (1926)

$\left[ 4\right] $ V. Bargmann, On a Hilbert space of analytic functions and
an associated integral transform, Part I, Commun. Pure Appl. Math. 14
(1961), 174-187

$\left[ 5\right] $J.P. Gazeau, J.R. Klauder, J. Phys. A: Math. Gen., 32
(1999) 123

$\left[ 6\right] $ Dodonov V V, 'Noncalssical' states in quantum optics: a
'squeezed review of the first 75 years, J.opt.B: Quantum Semiclass.opt. 4,
R1-R33 (2002)

$\left[ 7\right] $ J. P. Gazeau. Coherent States in Quantum Physics. Wiley,
Weinheim (2009).

$\left[ 8\right] $ H. A. Yamani, Z. Mouayn, Properties of shape invariant
tridiagonal Hamiltonians, Theor. Math. Phys. 203 (3) 761-779 (2020

$\left[ 9\right] $ Mourad E.H. Ismail, Classical and Quantum Orthogonal
Polynomials in one variable, Encyclopedia of Mathematics and its
applications, Cambridge university press (2005)

$\left[ 10\right] $ Koekoek, R., Lesky, P. A., and Swarttouw, R. F.
Hypergeometric orthogonal polynomials and their q-analogues. Springer
Monographs in Mathematics. Springer-Verlag, Berlin, 2010. With a \ \ \
foreword by Tom H. Koornwinder

$\left[ 11\right] $ G. Szego. Orthogonal polynomials. American Mathematical
Society Province, Rhode Island, 1939.

$\left[ 12\right] $ It\^{o} K., Complex multiple Wiener integral, Jap. J.
Math., 22 (1952) 63-86

$\left[ 13\right] $ Mourad E. H. Ismail, Analytic properties of complex
Hermite polynomials, Trans. Amer. Math. Soc., 2015

$\left[ 14\right] $ G. B. Folland, Harmonic Analysis in Phase Space, Ann.
Math. Stud., Vol.122, Princeton Univ. Press, Princeton, N.J. 1989.

$\left[ 15\right] $ M. E. Taylor, Noncommutative Harmonic Analysis, Math.
Surv. Monogr., 22, Amer, Math. Soc., Providence, R. I. (1989)

$\left[ 16\right] $ M. Duflo and C. C. Moore, J. Funct. Anal., 21, 209-243
(1976)

$\left[ 17\right] $ A. M. Perelomov, Generalized Coherent States and Their
Applications, 1986,\ Springer,\ Texts and Monographs in Physics 6, 156-164
(1971)

$\left[ 18\right] $ L. D. Abreu and H. G. Feichtinger, Functions spaces of
polyanalytic functions, Harmonic and Complex Analysis and its Applications,
Trends in Mathematics A. Vasilev (Ed.), Springer (2014) 139.

$\left[ 19\right] $ Z. Mouayn, Coherent states quantization for generalized
Bargmann spaces with formulae for their attacjed Berezin transform of the
Laplacian on $C^{n},$ J. Fourier Anal. App, 2012

$\left[ 20\right] $ Aronszajn N, Theory of reproducing kernels, Trans. Am.
Math. Soc.\ 68, pp.337-404 (1950)

$\left[ 21\right] $ Yamani, H.A., Mouayn, Z., Supersymmetry of the Morse
Oscillator (2016) Rep. Math. Phys. , 78 (3) 281-294

$\left[ 22\right] $ Yamani, H.A., Mouayn, Z., Coherent states associated
with tridiagonal Hamiltonians, Rep. Math. Phys. , 92 (I), pp.117-134 (2023)

$\left[ 23\right] $ P. Flandrin, Separability, positivity, and minimum in
time-frequency energy distribution, J. Math. Phys. 39 (8) 4016-4040 (1998)

$\left[ 24\right] $ D. Popov, Considerations concerning the harmonic limit
of the Morse oscillator, Physica Scripta, (63) \ 257-262 \ (2001)

$\left[ 25\right] $ H. Buchholz, The Confluent Hypergeometric Function, Vol
15, Springer-Verlag Berlin Heildelberg 1969.

$\left[ 26\right] $ A.P. Prudnikov, Yu. A. Brychkov and O.I. Marichev.
Integrals and Series, More special functions, vol 1, Gordon and Breach, 1986.

$\left[ 27\right] $ A.P. Prudnikov, Yu. A. Brychkov and O.I. Marichev.
Integrals and Series, More special functions, vol 2, Gordon and Breach, 1986.

$\left[ 28\right] $ Palam\`{a}, G., Sulla soluzione polynomiale della $%
\left( a_{0}+a_{1}x\right) y^{\prime \prime }+\left( b_{0}+b_{1}x\right)
y^{\prime }-nb_{1}y=0,$ Union Mathematica Italiana Bolletino,\ 1, 27-35
(1939)
\end{quote}

\end{document}